\documentclass[aps,showpacs,showkeys,nofootinbib,preprint]{revtex4}
\usepackage{epsfig,amssymb,bm,graphics,color,amsmath}

\begin{document} {\normalsize}

\title{Dynamical freeze-out phenomena: The case of 
$\mathbf K^\pm$, {\boldmath$\phi$}
transverse momentum spectra in collisions of
Au(1.23 A GeV) + Au}
%

\author{B.~Rabe  and B.~K\"ampfer}
 \affiliation{
 Helmholtz-Zentrum  Dresden-Rossendorf, 01314 Dresden, Germany}
 \affiliation{
 Institut f\"ur Theoretische Physik, TU~Dresden, 01062 Dresden, Germany}

\begin{abstract}
We argue on a continuous (dynamical) kinetic freeze-out of $K^\pm, \phi$ observed at
midrapidity in collisions Au(1.23 A GeV) + Au. The simulations by means 
of a transport model of BUU type point to time independent transverse
momentum slope parameters after 20 fm/c. The complex interplay of
expansion dynamics and strangeness production/exchange/absorption
as well as elastic scatterings involved in the reaction network does not 
support the previous interpretation of a late freeze-out of $K^-$ due
to larger cross sections.     
\end{abstract}

\pacs{25.75.-q; 25.75.Dw; 25.75.Ld}
\keywords{sub-threshold strangeness production, heavy-ion collisions, freeze-out}

\date{\today}

\maketitle

\section{Introduction}

The KaoS Collaboration \cite{Forster:2003vc} 
parameterized the transverse momentum spectra
of $K^\pm$, measured in collisions of Au(1.5 A GeV) + Au
at midrapidity, by means of J\"uttner type phase space distributions.
Ignoring the flow, the distributions become Boltzmann-like, depending essentially
on a slope parameter which is often coined "kinetic freeze-out temperature", $T$.
The experimental fact of $T_{K^+} > T_{K^-}$ has been interpreted 
in  \cite{Forster:2003vc} as
evidence for an expanding and cooling fireball with anti-kaons ($K^-$) decoupling 
at a later - thus cooler - stage due to their significantly larger cross section.
The kaons ($K^+$) decouple earlier - thus at a  hotter - stage.

An alternative interpretation has been put forward by the HADES Collaboration
\cite{Adamczewski-Musch:2017rtf}
for the reaction Au(1.23 A GeV) + Au: $K^+$ and $K^-$ can decouple under 
the same circumstances, i.e.\ at the same fireball temperature, thus 
$T_{K^+} = T_{K^-}$ at kinetic decoupling where the elastic interactions cease.
However, the unexpectedly large yield of $\phi$ changes the late spectra due to
the decays $\phi \to K^+ K^-$. Since $N_{K^+} \gg N_{K^-}$, the impact of
the $\phi$ decays on $T_{K^+}$ is minor. As observed in simulations \cite{Kotte_BK}, 
$T_{K^- from \, \phi \, decays}$ is approximately $\frac23 T_\phi$ for
distributions with $T_{K^\pm} \approx T_\phi$ prior to $\phi$ decays. Due to
the $26 \pm 8$ \% contribution of $K^-$ from $\phi$ decays, 
the observed final slope parameters
obey in fact $T_{K^+} > T_{K^-}$. In order to cope with the experimental 
results, a sufficiently large yield of $\phi$ is a prerequisite for such an interpretation
\cite{Adamczewski-Musch:2017rtf}.
(For an early assessment of relating $K^-$ and $\phi$ yields, 
see \cite{Kampfer:2001mc}, and for the first data-based quantitative analysis
of $\phi \to K^+ K^-$ feed-down, cf.\  \cite{Lorenz:2010zz}. 
For a comprehensive survey on strangeness
production in near-threshold heavy-ion and proton-nucleus collisions,
cf.~\cite{Hartnack:2011cn}, and for new simulation tools, see
\cite{Steinberg:2018jvv,Weil:2016zrk,Inghirami:2019muf,Steinheimer:2015sha}.)

Inspired by this controversy of the KaoS and HADES interpretations we performed
kinetic theory simulations of the collisions Au(1.23 A GeV) + Au and studied the
time evolution of the slope parameters $T_{K^\pm, \phi}$. Our findings can
be summarized as follows: $T_{K^\pm, \phi} (t > 20 \, \mbox{fm/c}) \approx const$.
(It should be stressed that the slope parameters $T_{K^\pm, \phi}$ refer to
transverse momentum spectra at midrapidity of all respective hadron species
$K^\pm$ and $\phi$; they are not related to a local or global fireball temperatures
steered overwhelmingly by nucleons and their excitations.) We interpret such a 
behavior as dynamic freeze-out. The energy dependence of the various inelastic
and elastic cross sections combined with the proper dilution upon expansion
of the fireball give rise to the conspiracy of 
$T_{K^\pm, \phi} (t > 20 \, \mbox{fm/c}) \approx const$. 
There is neither the need nor the possibility to define in an unambiguous manner
the freeze-out temperature at a certain instant of time within such a kinetic theory
approach.

The dynamical freeze-out phenomenon is familiar since some time for the
primordial nucleosynthesis in the early universe, where it refers however to the
abundances of light isotopes. Due to the energy dependence of cross sections,
which translate into a temperature dependence of reaction rates, combined with
the temporal temperature dependence due to the cosmic expansion, the isotopic
abundances $Y_i$ stay constant after about thousand seconds world age:
$Y_i (t > 10 \, \mbox{min}) \approx const$. While the starting
values are determined by statistical nuclear equilibrium \cite{Kolb_Turner}, 
the final values
are non-Markovian, i.e.\ depend on intermediate stages, and are very specific 
for cross sections (and some other parameters of the system). In particular,
the multitude of final (late) abundances can not be related to chemical equilibrium 
values at a certain common temperature.

A prototypical freeze-out model is provided in paragraph 5.2 in \cite{Kolb_Turner},
see Fig.~5.1 there. Based on the momentum-integrated Boltzmann equation
the normalized abundance of a massive-particle species is seen to follow for
some time the fiducial equilibrium abundance but levels off at certain point
and stays constant afterwards, never reaching the ever-dropping equilibrium
value. The leveling off is determined by a combination of the expansion (cooling) rate and
the reaction rate based on thermally averaged cross section. 
That is, at and after freeze-out, the expansion rate exceeds the reaction rate.  

From such a perspective, it is astonishing that the various hadron abundances 
in heavy-ion collisions can be described by chemical equilibrium values 
at a common (albeit beam-energy dependent)
temperature modulo a normalization volume. At LHC energies, no
other parameters are needed to uncover the hadron and isotopic (including
anti-nuclei) yields over nine orders of magnitudes \cite{Andronic:2017pug}; 
at lower beam energies,
the baryo-chemical potential becomes important, which is also beam-energy
dependent, to describe a multitude of hadron yields, 
cf.~\cite{Agakishiev:2015bwu,Agakishiev:2010rs}
for examples.

Besides the chemical freeze-out, related to abundances, also the kinetic freeze-out,
related to the momentum distributions of hadrons, is of interest as a signature
of the "hadronic life", e.g.\ after the hadronization of deconfined strong-interaction 
matter at LHC energies. This freeze-out dynamics may be flavor dependent
and may be different for ground state hadrons and short-living resonances
\cite{Motornenko:2019jha}. In the fragmentation region, large net-baryon
densities are expected in ultra-relativistic heavy-ion collisions \cite{Kapusta:2018omb} -
quite similar to conditions achieved in relativistic heavy-ions. 
Using rare (i.e.\ strange and charm) probes of strongly compressed
baryon matter is a central part of the research program of the CBM
collaboration \cite{Friman:2011zz,Ablyazimov:2017guv}.
These facets of ultra-relativistic
heavy-ion collisions in turn are linked to medium-energy (relativistic)
heavy-ion collisions, where one gains complementary important information
on hadronic many-body dynamics.  
              
After this digression on freeze-out phenomena, let us return to the primary goal
of our study - the time evolution of $K^\pm, \phi$ slope parameters in a kinetic
theory simulation of BUU type for the reaction Au(1.23 A GeV) + Au. The BUU
model is briefly described in section~\ref{sec:buu},
and its numerical results are spelled out in section~\ref{numerical_results}.
Since we employ a code version which has been
successfully utilized in \cite{Schade:2009gg} for the reaction Ar(1.75 A GeV) + KCl, 
we present 
analog results of the time evolution of $T_{K^\pm, \phi}$ in Appendix~A  
for a comparison. 
We summarize in section~\ref{sec:summary}.

\section{BUU code}\label{sec:buu}
         
\begin{figure}[htb!]
\centering
\begin{minipage}[b]{0.495\textwidth}
	\includegraphics{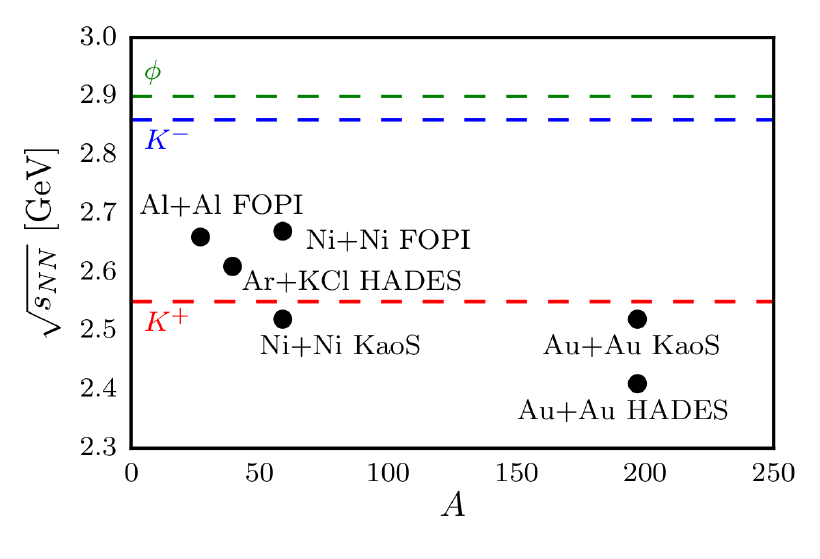}
\end{minipage}
\begin{minipage}[b]{0.495\textwidth}
      \includegraphics{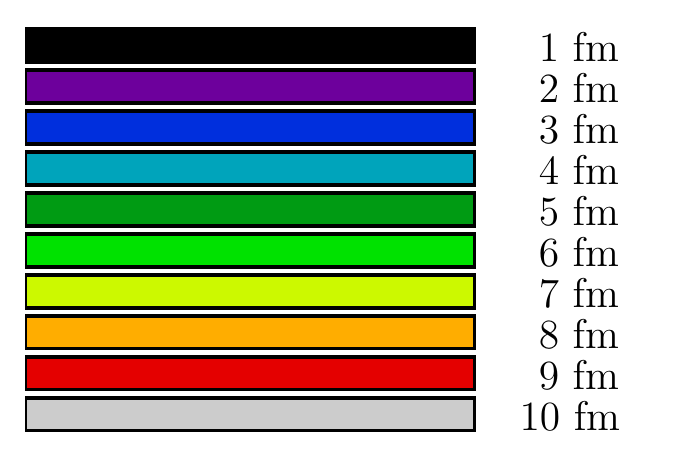}
\vspace*{9mm}
\end{minipage}
\vspace*{-9mm}
 \caption{Left panel: Overview of the center-of-mass energies for various experiments \cite{Adamczewski-Musch:2017rtf, Forster:2003vc, Agakishiev:2009ar, Gasik:2015zwm, Piasecki:2018psj} (bullets) and the free \(NN\) threshold energies of 
\( K^\pm \) and \( \phi \) mesons (dashed lines).
Right panel: Color code used for the various impact parameters.}
 \label{fig:2.1}
 \end{figure}

We employ here the same BUU code as utilized in \cite{Schade:2009gg} 
for collisions Ar(1.75 A GeV) + KCl.
Only the beam energy, system size, proton-to-neutron ratio, and
impact parameter range are adopted; all other parameters are frozen in,
see Tab.~\ref{tab:3.1}. The code was compared in \cite{Kolomeitsev:2004np} 
with other transport
codes and some peculiarities have been identified. Nevertheless, as shown
in \cite{Schade:2009gg} the BUU code copes successfully with the data 
\cite{Agakishiev:2009ar}. The decisive difference
of the reactions Au(1.23 A GeV) + Au and Ar(1.75 A GeV) + KCl are the significantly
lower beam energy and the significantly larger system size of the former one.
This lets us expect a larger sensitivity to many-body effects, in particular in-medium
effects, since the reactions with strange mesons involved are deeper below the 
respective thresholds. Figure~\ref{fig:2.1}, left panel,  
exhibits a survey on the thresholds
and experiments performed up to now in the threshold region.

\begin{table}[h]
	\begin{tabular}{|l|c|c|c|}
		\hline
		Input parameter & Symbol & Au+Au & Ar+KCl \\
		\hline
		Simulation duration & \(t_{\text{max}}\) [fm/c] & 60 & 60 \\
		Time step & \(\delta t\) [fm/c] & 0.5 & 0.5 \\
		Projectile & \(A_P\) / \(Z_P\) & 197 / 79 & 40 / 18 \\
		Target & \(A_T\) / \(Z_T\) & 197 / 79 & 39 / 19 \\
		Kinetic energy & \(E_{\text{kin}}\) [AGeV] & 1.23 & 1.756 \\
		Initial distance between nuclei & \(r_{\text{dist}}\) [fm] & 2.9 & 2.9 \\
		Impact parameter & \(b\) [fm] & 1-10 & 1-6 \\
		Number of parallel ensembles & \(\tilde{N}\) & 200 & 200 \\
		Number of subsequent iterations & isubs & 200 & 200 \\
		Incompressibility & \(\kappa\) [MeV] & 215 & 215 \\
		Effective mass shift \(K^+\) & \(\Delta m_{K^+}(\rho_0)\) [MeV] & \(+23.5\) & \(+23.5\) \\
		Effective mass shift \(K^-\) & \(\Delta m_{K^-}(\rho_0)\) [MeV] & \(-75.2\) & \(-75.2\) \\
		Effective mass shift \(\phi\) & \(\Delta m_{\phi}(\rho_0)\) [MeV] & \(-22.2\) & \(-22.2\) \\
		Nuclear saturation density & \(\rho_0\) [\(\mathrm{fm^{-3}}\)] & 0.16 & 0.16 \\
		\hline
	\end{tabular}
\caption{Several input parameters for the simulations of 
collisions Au+Au and Ar+KCl.}
\label{tab:3.1}
\end{table}  

\section{Numerical results for A\lowercase{u}(1.23 A GeV) + A\lowercase{u}}\label{numerical_results}

\subsection{Impact parameter dependence}

\begin{figure}[htb!]
	\centering
	\begin{minipage}[t]{0.495\textwidth}
		\includegraphics{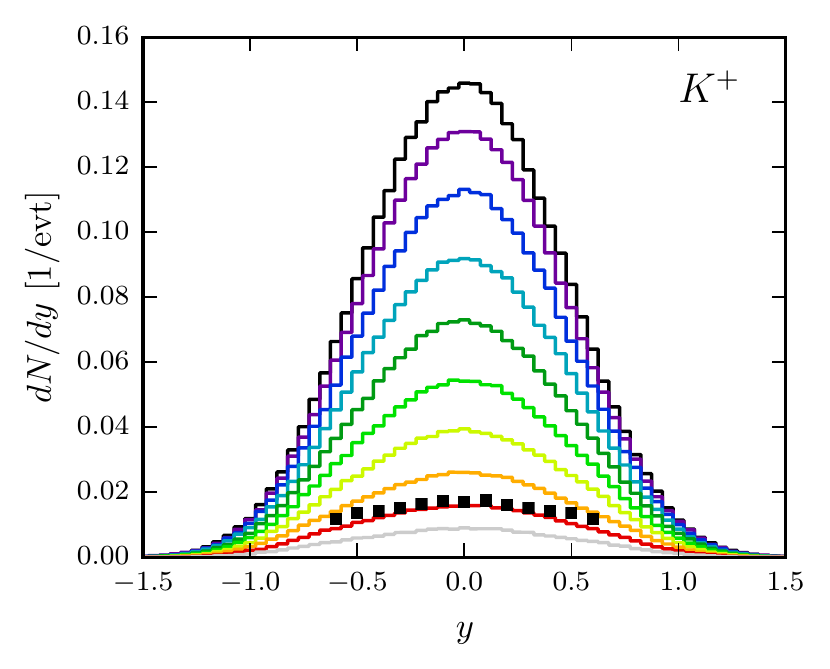}
	\end{minipage}
	\begin{minipage}[t]{0.495\textwidth}
		\includegraphics{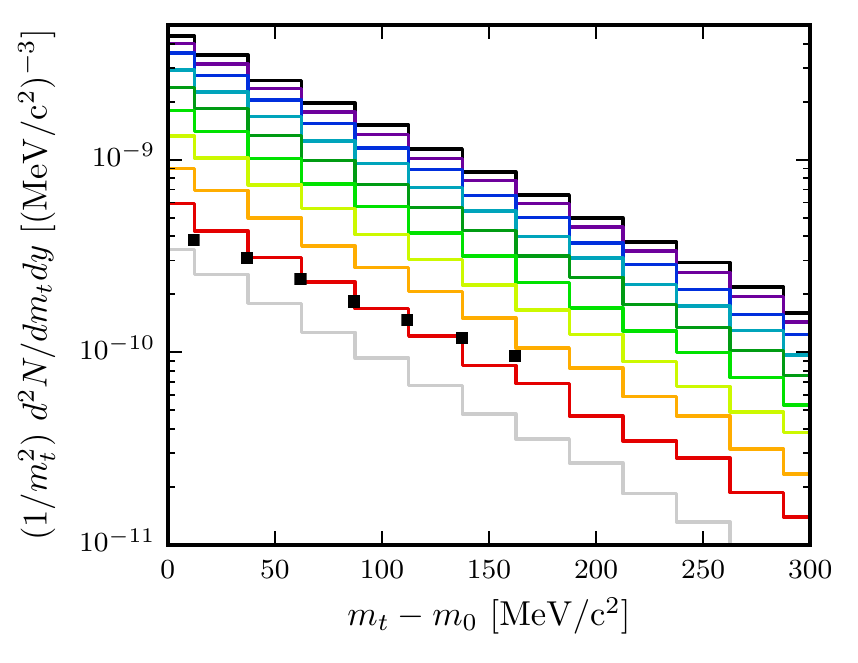}
	\end{minipage}
\vspace*{-9mm}
	\caption{Left panel: Rapidity spectra of \( K^+ \) mesons in the center-of-mass system for impact parameters \( 1 \leq b \leq 10 \)~fm (top to bottom). Right panel: Transverse mass spectra within the rapidity interval \( -0.1 \leq y \leq 0.1 \)
for the same impact parameter range as in left panel. 
The black symbols represent the experimental data from 
\cite{Adamczewski-Musch:2017rtf} (centrality 20 -40 \%). 
The color code for the impact parameters is depicted in the right panel of Fig.~\ref{fig:2.1}.
}
 \label{fig:3.1}
 \end{figure}

\begin{figure}[htb!]
	\centering
	\begin{minipage}[t]{0.5\textwidth}
		\includegraphics{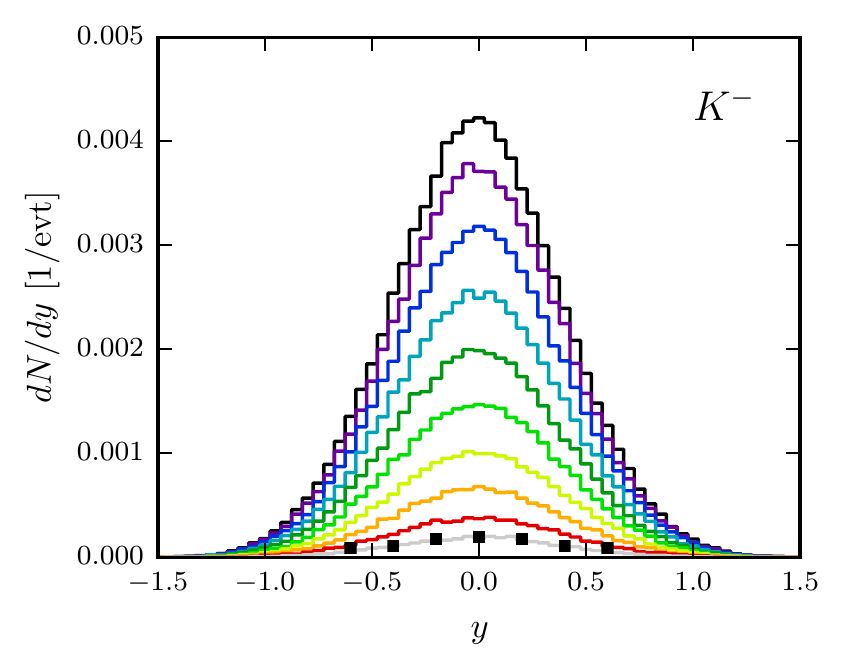}
	\end{minipage}
	\begin{minipage}[t]{0.49\textwidth}
		\includegraphics{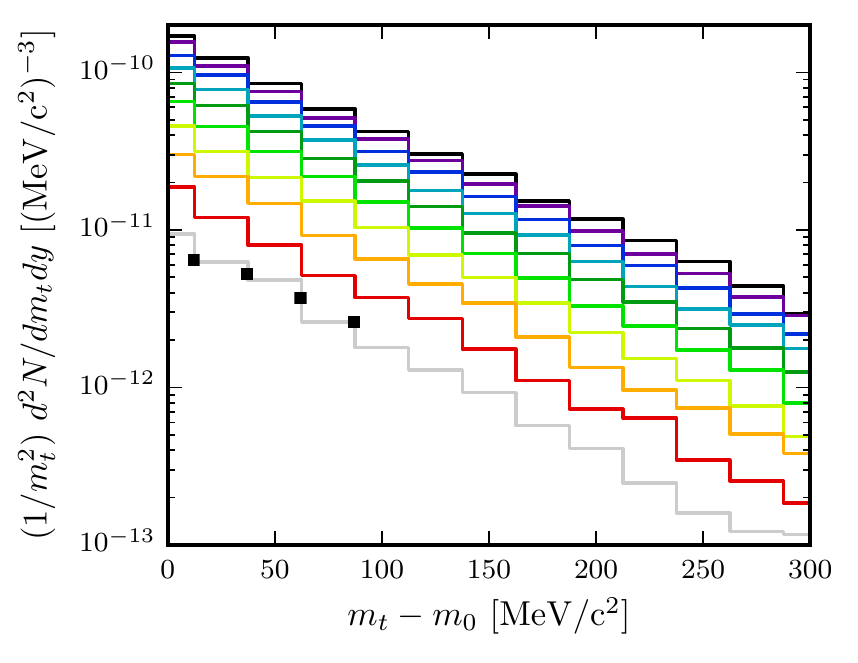}
	\end{minipage}
\vspace*{-9mm}
 \caption{Same as Fig.~\ref{fig:3.1}, but for \(K^-\) mesons.}
 \label{fig:3.2}
 \end{figure}

\begin{figure}[htb!]
	\centering
	\begin{minipage}[t]{0.51\textwidth}
		\includegraphics{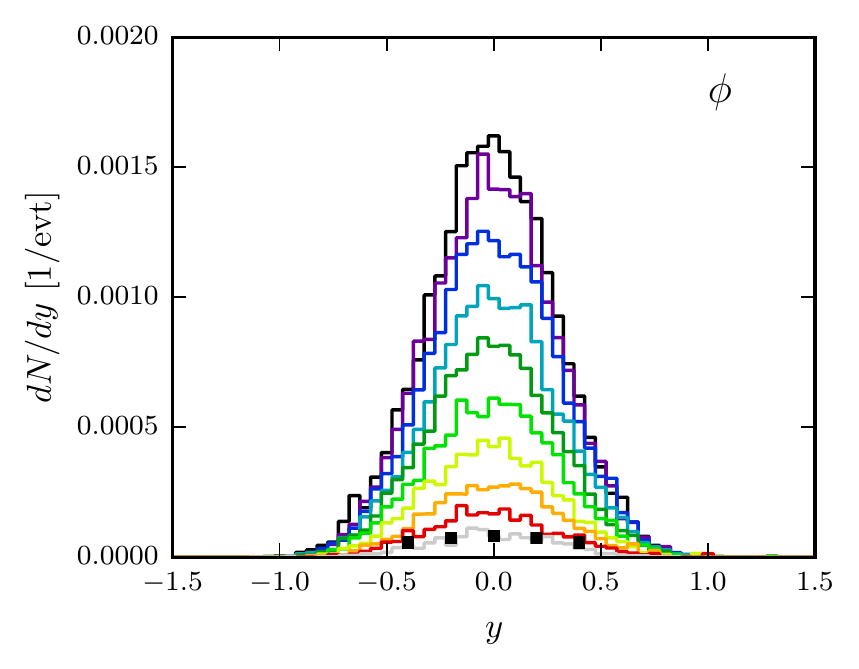}
	\end{minipage}
	\begin{minipage}[t]{0.48\textwidth}
		\includegraphics{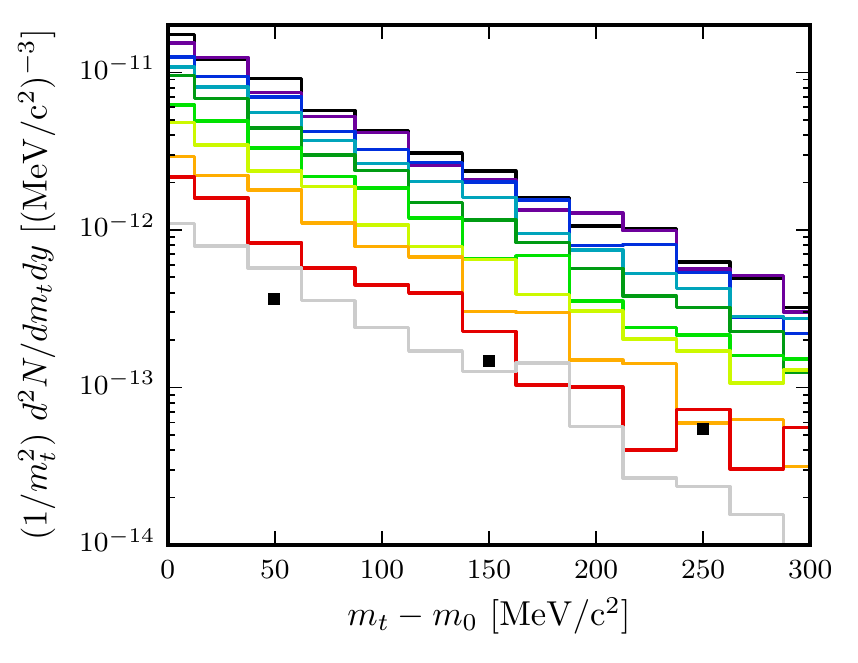}
	\end{minipage}
\vspace*{-9mm}
	\caption{Same as Fig.~\ref{fig:3.1}, but for \(\phi\) mesons.}
 \label{fig:3.3}
 \end{figure}

It happened that $K^\pm,\phi$ transverse momentum spectra and rapidity
distributions of the Ar(1.75 A GeV) + KCl data \cite{Agakishiev:2009ar} 
could be described very well
by the impact parameter $b = 3.9$~fm, see Appendix A. 
Following such a strategy for 
Au(1.23 A GeV) + Au we see that an optimum description of the data 
\cite{Adamczewski-Musch:2017rtf} for the centrality class 0 - 40 \%
is accomplished by $b = 9$~fm for $K^+$ and $b = 10$~fm for $K^-, \phi$,
see Figs.~\ref{fig:3.1}, \ref{fig:3.2} and \ref{fig:3.3}. As a compromise we
use henceforth $b = 9$~fm to avoid subtleties of certain impact parameter
averaging procedures according to weighting $\propto b \, db$. In fact, Table II in 
\cite{Adamczewski-Musch:2017sdk} attributes the 0 - 40 \% centrality class
to the impact parameter interval $b = 0 - 9.3$~fm with a mean of 6.2~fm.
The upper panel in Fig.~8 of \cite{Adamczewski-Musch:2017sdk} indicates that
an impact parameter range 7 - 11~fm centered at 9~fm corresponds to the centrality
class 30 - 40 \%. (Table II in \cite{Adamczewski-Musch:2017sdk}
quotes a mean impact parameter of 8.71~fm for that centrality class.)
Therefore, the selection of one "representative impact parameter"
must be considered with caution. To get some feeling on the impact parameter
dependence we exhibit in the following often the sequence of $b = 1 - 10$~fm
in steps of 1~fm with color code displayed in the right panel of Fig.~1.  

The transverse momentum spectra unravel some deficits of our model:
$T_{K^\pm}$ are somewhat too low, while $T_\phi$ is much too low,
see Fig.~10 below. For $\phi$,
these deficits can not be cured by some impact parameter averaging since
we fail to meet the experimental values of $T_{\phi}$ for all values of $b$. 
On the other hand,  \cite{Adamczewski-Musch:2017rtf} quotes slope
parameters of about 91~MeV ($K^+$, centrality class 30 - 40 \%)
and $69 \pm 7$~MeV ($K^-$, centrality class 20 - 40 \%), which are not
too small in comparison with our results quantified in Fig.~10 below.  
As in experiment \cite{Adamczewski-Musch:2017rtf} we define the slope parameters by
\begin{equation}\label{eq:3.2}
\frac{1}{m_t^2} \frac{d^2 N}{dm_t dy} \bigg|_{y_0} = C(y_0) 
\exp \bigg(- \frac{(m_t - m_0)}{T_B(y_0)}\bigg)
\end{equation}
at midrapidity $y_0$ in the rapidity interval $y = y_0 \pm 0.1$. 
In addition, a normalization
factor $C$ is attributed separately to each species. The transverse mass is defined by
$m_t^2 = p_t^2 + m_0^2$ with rest masses $m_0$ of the considered species.
Henceforth, we denote $T_B (y_0) = T$ which is again specific for each specie.

\subsection{Nucleon density evolution}

\begin{figure}[htb!]
	\centering
	\includegraphics{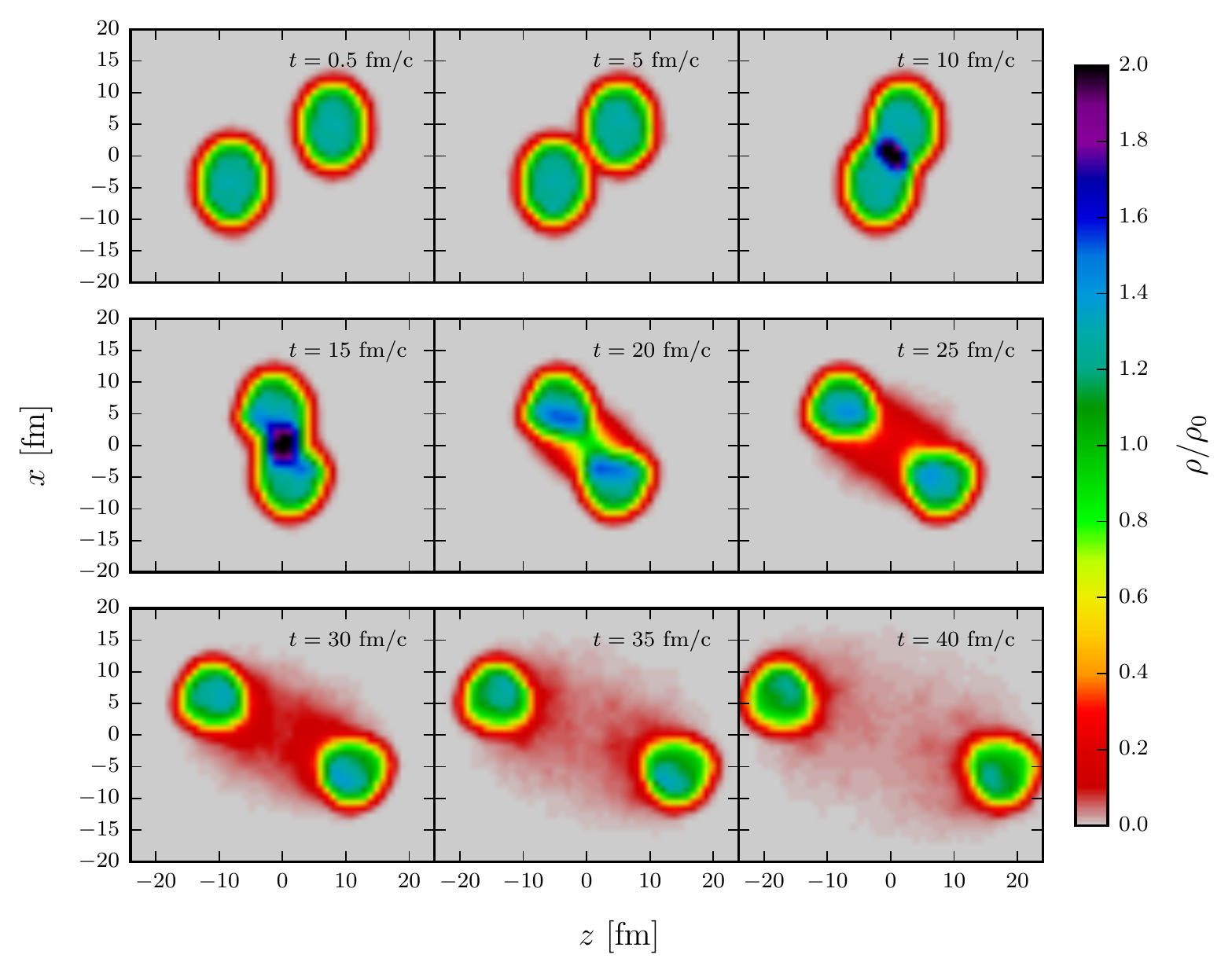}
\vspace*{-3mm}
	\caption{Snapshots of the projection of the normalized nucleon density 
in the reaction plane for the collision Au(1.23 A GeV) + Au for the impact parameter 
\( b = 9 \)~fm at times depicted in the legends.}
 \label{fig:3.6}
 \end{figure}  

\begin{figure}[htb!]
	\centering
	\includegraphics{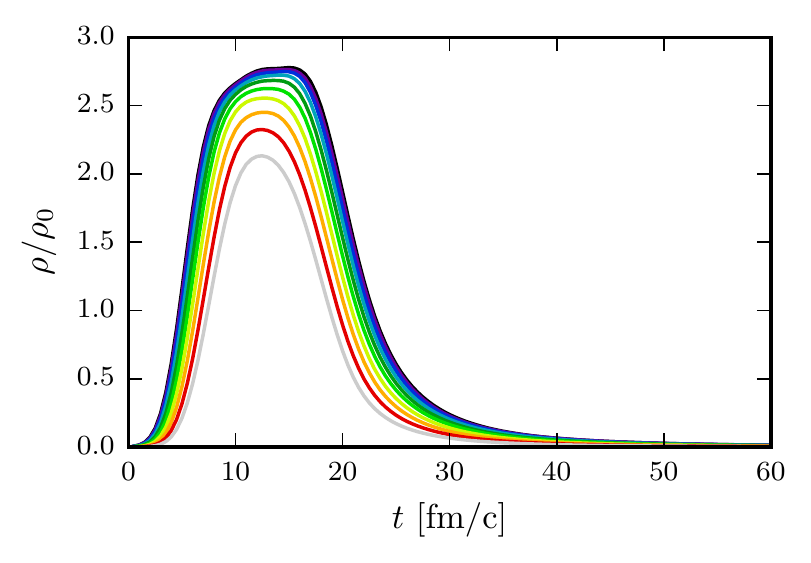}
\vspace*{-3mm}
	\caption{Normalized nucleon density in the central cell of volume
 \( 1 \ \mathrm{fm^3} \) as a function of time for the impact parameters
 \( 1 \leq b \leq 10 \)~fm.} 
 \label{fig:3.7}
 \end{figure}

We refrain here completely from determining any local temperature. Instead, to
visualize the time evolution of the system by the nucleon density in the reaction
plane, we exhibit in Fig.~\ref{fig:3.6} several snapshots for the representative
impact parameter $b = 9$ fm. The maximum nucleon density is about $2 \rho_0$
(or $2.75 \rho_0$ for $b = 1$ fm) at $t \approx 12$ fm/c in the central cell of
volume 1~fm${}^3$. The high-density stage with $\rho \ge\rho_0$ is for
$t = 7 \cdots 20$ fm/c, see Fig.~\ref{fig:3.7}     

\subsection{Time evolution of {\boldmath$K^\pm, \phi$} rates}

\begin{figure}[htb!]
	\centering
	\begin{minipage}[t]{0.495\textwidth}
		\includegraphics{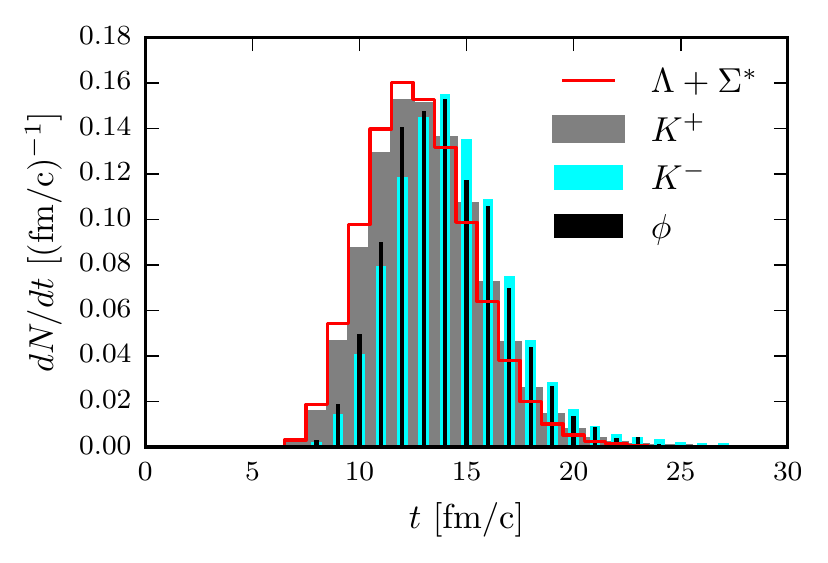}
	\end{minipage}
	\begin{minipage}[t]{0.495\textwidth}
		\includegraphics{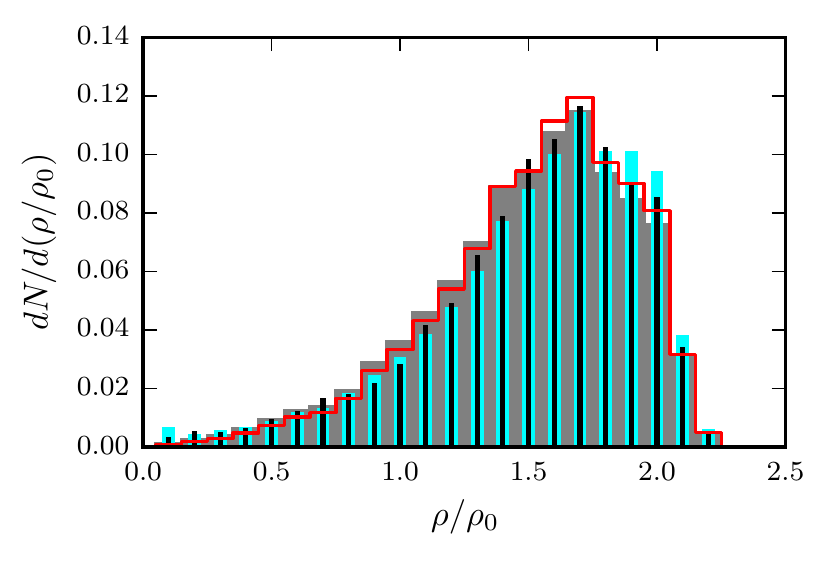}
	\end{minipage}
\vspace*{-9mm}
\caption{Normalized production rates of \(K^\pm\) and \(\phi\) mesons as well as \(\Lambda\) and \(\Sigma^*\) hyperons as functions of time (left) and local nucleon density (right) for \(b = 9\)~fm. The normalization was performed over the respective total multiplicities of productions up to final run time of 60 fm/c.}
 \label{fig:3.8}
 \end{figure}

\begin{figure}[htb!]
	\centering
	\begin{minipage}[t]{0.495\textwidth}
		\includegraphics{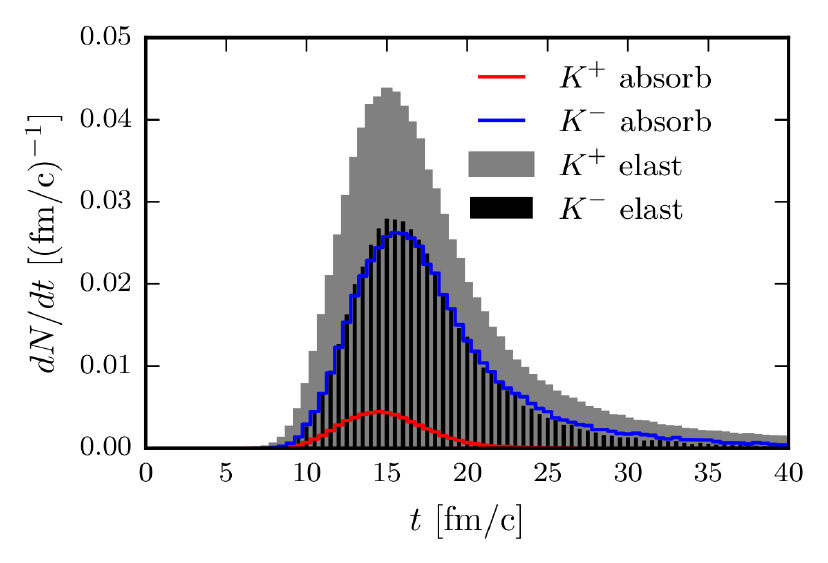}
	\end{minipage}
	\begin{minipage}[t]{0.495\textwidth}
		\includegraphics{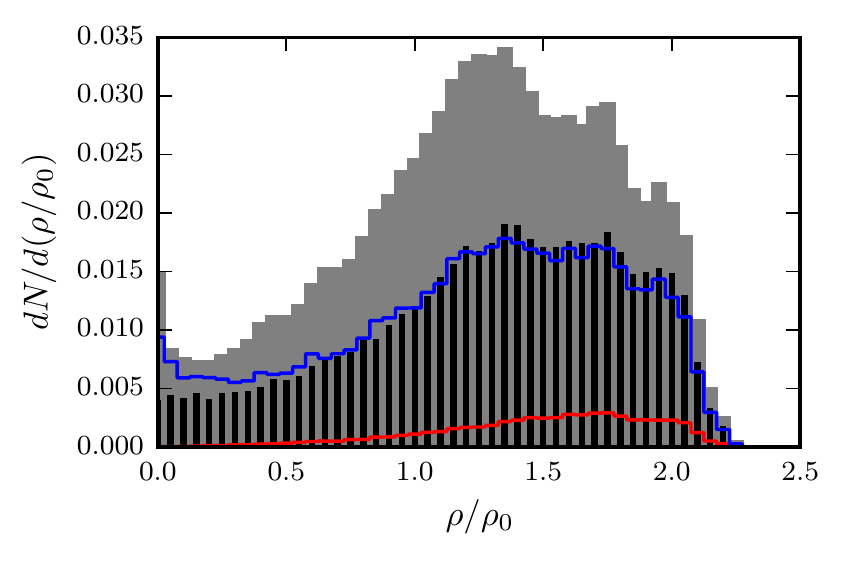}
	\end{minipage}
\vspace*{-9mm}
\caption{Normalized collision rates of \( K^\pm \) mesons with nuclear matter as functions of time (left) and local nucleon density (right) for \( b = 9 \)~fm. Normalization as in Fig.~\ref{fig:3.8}. 
}
 \label{fig:3.10}
 \end{figure}

\begin{figure}[htb!]
	\centering
	\begin{minipage}[t]{0.495\textwidth}
		\includegraphics{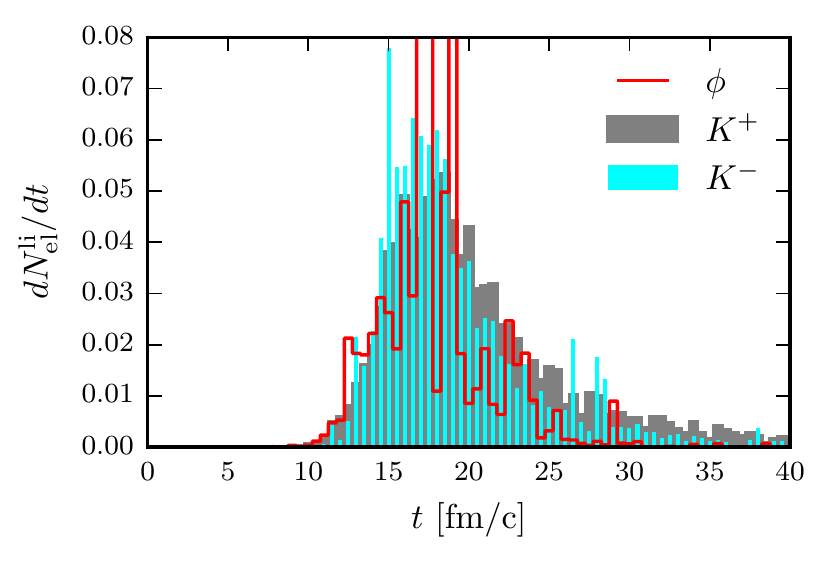}
	\end{minipage}
	\begin{minipage}[t]{0.495\textwidth}
		\includegraphics{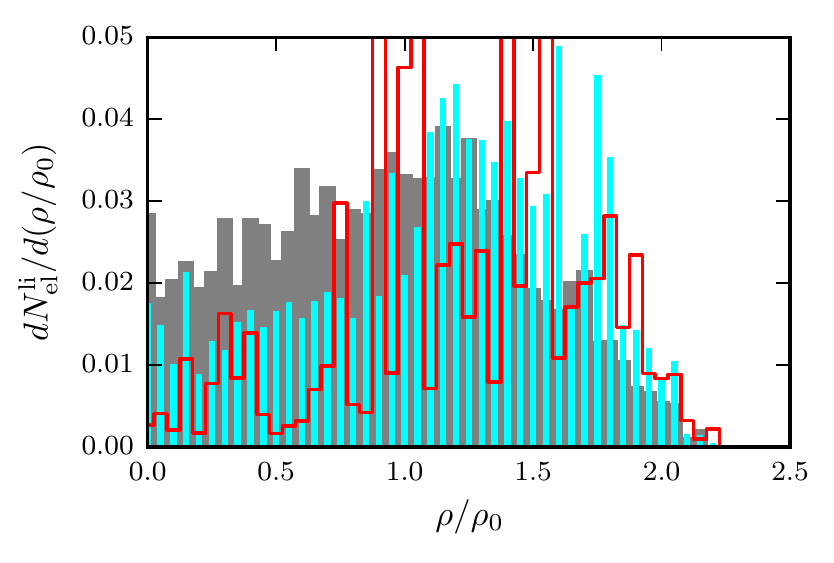}
	\end{minipage}
\vspace*{-9mm}
\caption{Normalized rates of last elastic interactions 
of \( K^\pm \) and \( \phi \) mesons with nuclear matter as functions of time (left) and local nuclear density (right) for \( b = 9 \)~fm. Normalization as in Fig.~\ref{fig:3.8}. 
}
 \label{fig:3.11}
 \end{figure}

Production rates of $K^\pm, \phi$ as well as $\Lambda + \Sigma^{\pm, 0}$
as a function of time (left panel) 
are displayed in Fig.~\ref{fig:3.8}. The analog elastic collision rates of 
$K^\pm$ with nucleons and the absorption rates are exhibited in Fig.~\ref{fig:3.10}.
Concerning the production rates one sees some delay of $K^-, \phi$ relative to
$K^+, \Lambda + \Sigma^{\pm, 0}$.
For brevity we denote by $\Sigma^*$ all $\Sigma^{\pm, 0}$.
Focusing still on the evolution one can not see a later decoupling of $K^-$;
instead the last elastic interactions of $K^+$ seem to go on for a somewhat longer
time, see Fig.~\ref{fig:3.11}. These investigations are aimed at elucidating
whether there is a clear time ordering of production and freeze-out
of $K^\pm$ and $\phi$. 

Since the in-medium masses of $K^\pm$ are strikingly different
(see Table I), it has been argued that production / rescattering / absorption /
last elastic interaction probe different densities. In fact, $K^-$ production
happens at somewhat
larger densities (see right panel of Fig.~\ref{fig:3.8}); nevertheless, the $K^\pm$
production rates peak at about $1.6 \rho_0$. Elastic rescatterings peak at $1.3 \rho_0$
($K^+$) and $1.4 \rho_0$ ($K^-$), respectively, while the absorptions acquire
maxima at $1.3 \rho_0$ ($K^-$) and $1.7 \rho_0$ ($K^+$), see right panel
in Fig.~\ref{fig:3.10}. The maximum 
of last interaction rates is at $1.2 \rho_0$ ($K^\pm$), however, with more
last interactions of $K^+$ at lower densities, $\rho < \rho_0$, see right panel
in Fig.~\ref{fig:3.11}. This seems to be in contrast to the above mentioned
time and density ordering of $K^+$ and $K^-$ collisions.  

\subsection{Time evolution of {\boldmath$T_{K^\pm, \phi}$ }  }

\begin{figure}[htb!]
	\centering
	\includegraphics{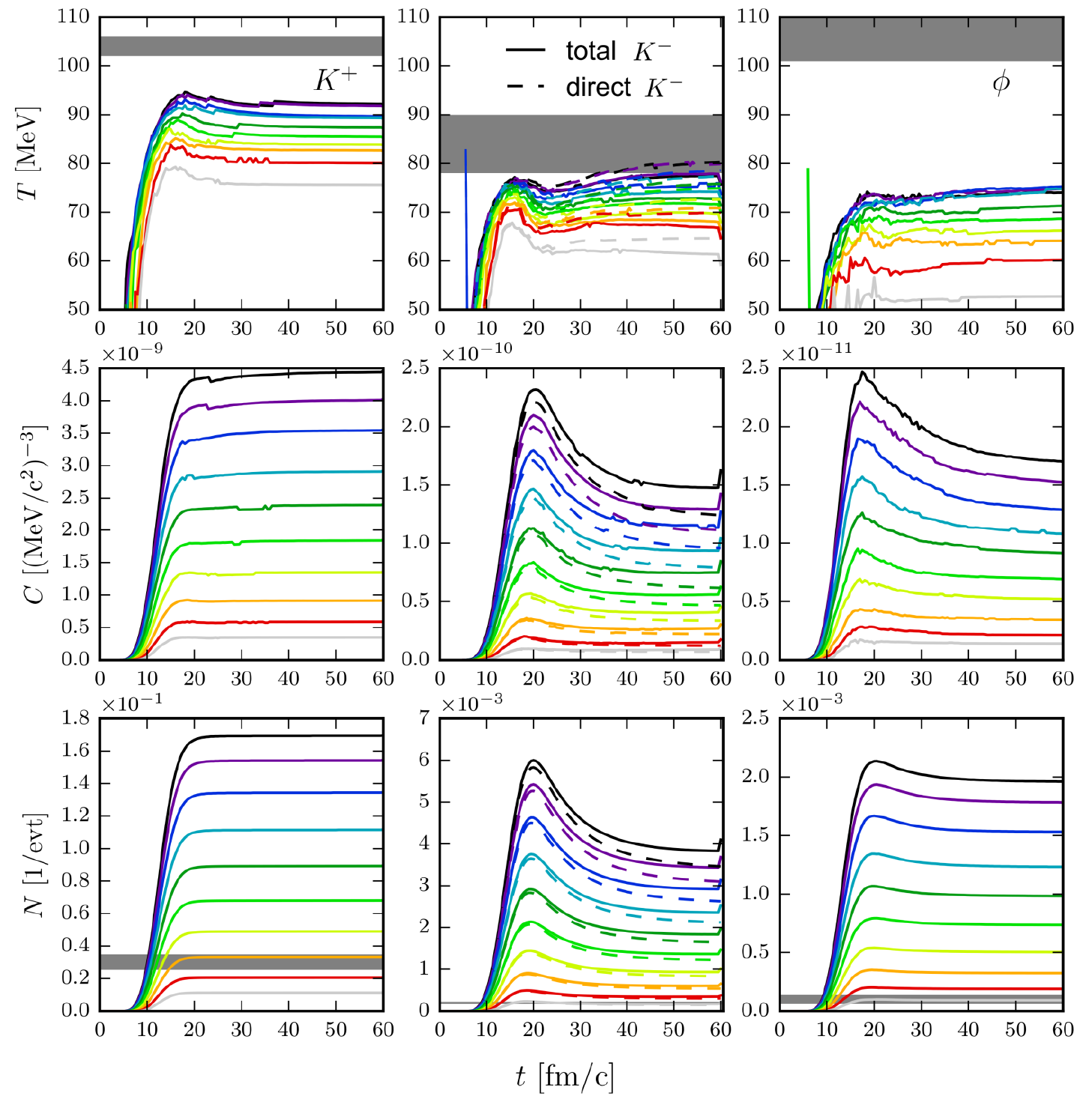}
\vspace*{-3mm}
	\caption{Effective temperature \( T \) parameterizing the
transverse momentum spectra (top row),  
normalization \( C \) (middle row)
and multiplicity \( N \) (bottom row)
of \( K^+ \) (left column), \(K^-\) (middle column) and \( \phi \) mesons (right column) as functions of time for impact parameters \( 1 \leq b \leq 10 \)~fm. 
The experimental data \cite{Adamczewski-Musch:2017rtf} for the centrality class
0 - 40 \% are represented as gray bars.
"total $K^-$" means inclusion of $K^-$ from $\phi$ decays, while
"direct $K^-$" is without that contribution.}
 \label{fig:3.12}
 \end{figure}  

The time evolution of the parameters $T_{K^\pm, \phi}$ and $C_{K^\pm, \phi}$
defined in Eq.~\eqref{eq:3.2} is exhibited in Fig.~\ref{fig:3.12} together with the
$K^\pm, \phi$ multiplicities. The most striking point is 
$T_{K^\pm, \phi} \approx const$ for $t > 20$ fm/c and all impact parameters;
$T_{K^-}$ is even slightly increasing after 20 fm/c for the most central collisions.
This invalidates the expectation that the slope parameter $T$ can be related to
a local medium temperature. Rather, the interaction rates (see Figs.~\ref{fig:3.10}
and \ref{fig:3.11}) drop to zero at $t > 40$ fm/c, meaning the dynamic freeze-out
in the same spirit as in the Big Bang Nucleosynthesis. Prior to 40 fm/c, but after 
20 fm/c, the intricate network of production, absorption and elastic reactions on top
of the overall expansion (see Figs.~\ref{fig:3.6} and \ref{fig:3.7}) result in 
$T_{K^\pm,\phi} \approx const$.

\begin{figure}[htb!]
	\centering
	\includegraphics{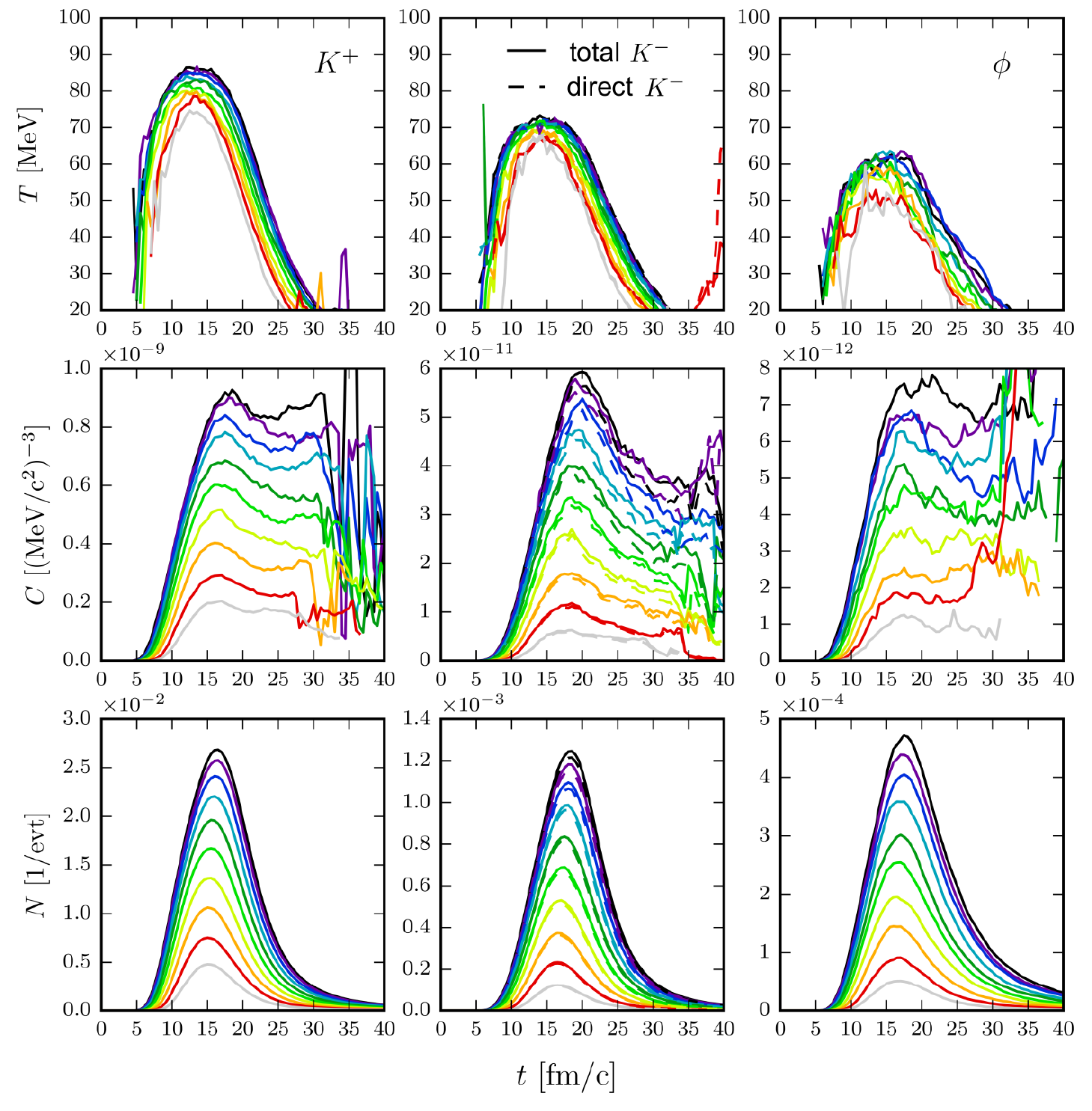}
\vspace*{-3mm}
	\caption{Same as Fig. \ref{fig:3.12}, but for mesons within the central cell of volume \( 2.5 \times 2.5 \times 2.5 \ \mathrm{fm^3} \).}
 \label{fig:3.14}
 \end{figure}  

\begin{figure}[htb!]
	\centering
	\includegraphics{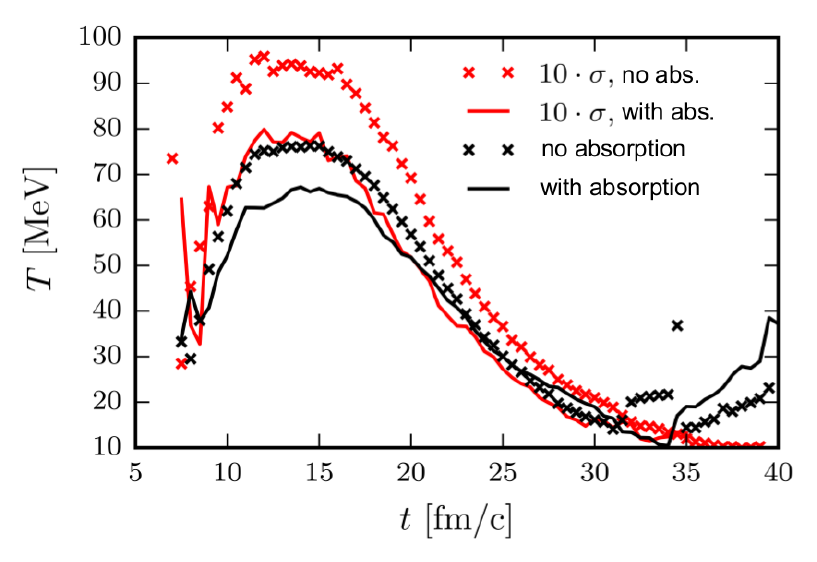}
\vspace*{-3mm}
	\caption{Influence of \( K^- \) absorption as well as up-scaling of the total cross section \( \sigma_{\text{tot}} \) by factor~\(10\) on the time dependence 
of the slope parameter, \( T_{K^-} (t) \).  
For \( b = 9 \)~fm, central cell.}
 \label{fig:3.16}
 \end{figure}  

A completely different picture emerges when considering the parameters
$T_{K^\pm,\phi}$ in the central cell: After 15 fm/c ($K^\pm$) or 17 fm/c
($\phi$), the values of  $T_{K^\pm,\phi}$ rapidly drop and become $< 20$ MeV
for $t > 30$ fm/c, see Fig.~\ref{fig:3.14}. Of course, such a correlation of momentum
space and position space is experimentally hardly accessible. We emphasize in this
context the importance of cross sections. For instance, scaling up the total 
$K^-+ anything$ cross section by a factor of ten increases the maximum of $T_{K^-}$
by 15 MeV; switching off the absorption channels lets the maximum of $T_{K^-}$ 
further increase towards 100 MeV, see Fig:~\ref{fig:3.16}. Switching off the
absorption for our standard setting of cross sections (see Appendix A in \cite{BRabe})
also increases the maximum of $T_{K^-}$ by about 15 MeV.
 
\begin{figure}[htb!]
	\centering
	\begin{minipage}[t]{0.495\textwidth}
		\includegraphics{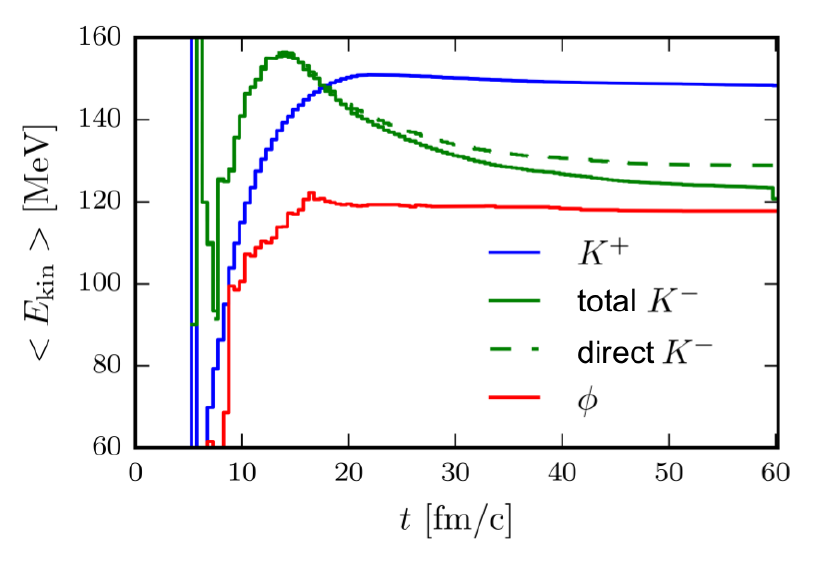}
	\end{minipage}
	\begin{minipage}[t]{0.495\textwidth}
		\includegraphics{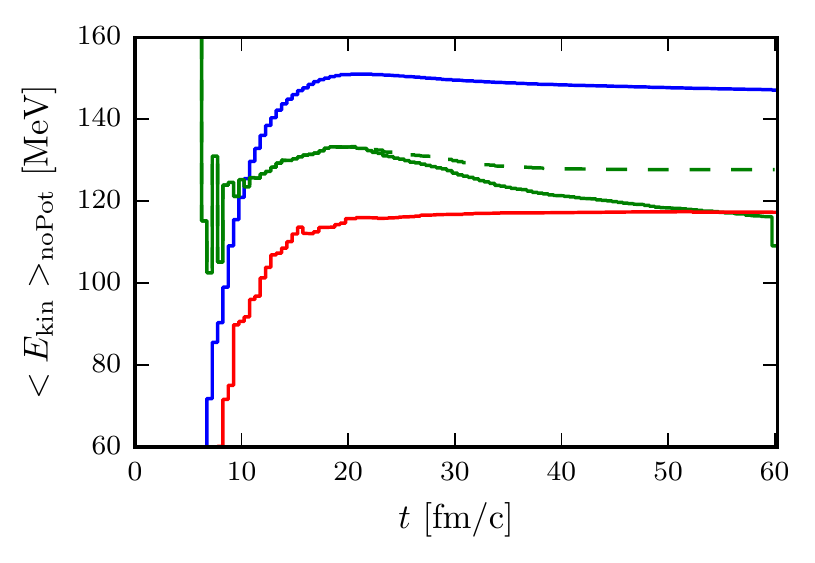}
	\end{minipage}
\vspace*{-9mm}
\caption{Mean kinetic energy with (left) and without \(KN\) potential 
(right) of \( K^\pm \) and \( \phi \) mesons as functions of time for \( b = 9 \)~fm.}
 \label{fig:3.18}
 \end{figure}  

A quantity, which can serve as a proxy of the temperature in local off-equilibrium 
situation, is the mean kinetic energy. We define $E_{kin} = E - m^*$
in the center-of-mass system. 
The effective in-medium mass is defined by $m^* = m + \Delta m \rho$
with values of $\Delta m \rho_0 = \Delta m(\rho_0)$ listed in Table I.
Interestingly, we observe 
$\langle E_{kin, K^-} \rangle > \langle E_{kin, K^+} \rangle$
at $t < 18$ fm/c, see Fig.~\ref{fig:3.18} - left panel.
The pronounced dropping of  $\langle E_{kin, K^-} \rangle$ at $t > 13$ fm/c can be attributed
to the change of the effective in-medium mass $m^*$ with dropping nucleon density.
In fact, switching off the effective in-medium masses, i.e.\ $m_* \to m_0$,
the dropping of $\langle E_{kin, K^-} \rangle$ ceases, see dashed green curve in right panel of Fig.~\ref{fig:3.18}. 
Since the in-medium modifications of $K^+$ and
$\phi$ are minor, there is no noticeable impact on 
$\langle E_{kin, K^+, \phi} \rangle$ vs. $t$.
The solid green curves in  Fig.~\ref{fig:3.18} include the $K^-$ from $\phi$ decays.
The "cooling" of the $K^-$ spectrum is due to the above mentioned fact that the
mean kinetic energies of decay-$K^-$ are less than the mean kinetic energies of
$K^-$ in a medium with a temperature scale ${\cal O} (100)$ MeV. 

\section{Discussion and summary}\label{sec:summary}
 
Contrary to the interpretations \cite{Forster:2003vc,Adamczewski-Musch:2017rtf} of  $T_{K^+} > T_{K^-}$ in relativistic
heavy-ion collisions at sub-threshold beam energy, the use of a BUU transport 
simulation points to a more complex picture. Figure~\ref{fig:3.12}
suggests that from the very beginning of the fireball expansion the relation
$T_{K^+} > T_{K^-}$ is established and does not change in the subsequent
evolution. That is, the complicated interplay of various reaction channels involved
in strangeness production, exchange, and absorption  up to elastic scatterings
cause flatter $K^+$ transverse momentum spectra at midrapidity. We emphasize
that the slope parameters $T_{K^\pm, \phi}$ are used to parameterize these
transverse momentum spectra. Enforcing a correlation to position space, e.g.\
by considering only mesons in the central cell, changes substantially the behavior
of $T_{K^\pm, \phi}$. Instead of staying approximately constant (as for
all mesons in the fireball), it rapidly drops after achieving a maximum. The picture
becomes
more obscured by analyzing the mean kinetic energies of $K^\pm,\phi$. The coupling of the effective in-medium mass to the nucleon density
causes a significant reduction of the  $K^-$ kinetic energy below the $K^+$
kinetic energy, while without in-medium effects the $K^-$ kinetic energy is
below the $K^+$ kinetic energy, both ones being roughly constant in time.

Irrespectively of the temporal aspects we tried also to check the hypothesis that the
$K^-$ have at the production instant lower kinetic energies due to peculiarities of the
production channels. However, subsequent absorption and scattering/transfer  
reactions make such hypothesis not convincingly enough.

\begin{figure}[htb!]
\centering
\includegraphics{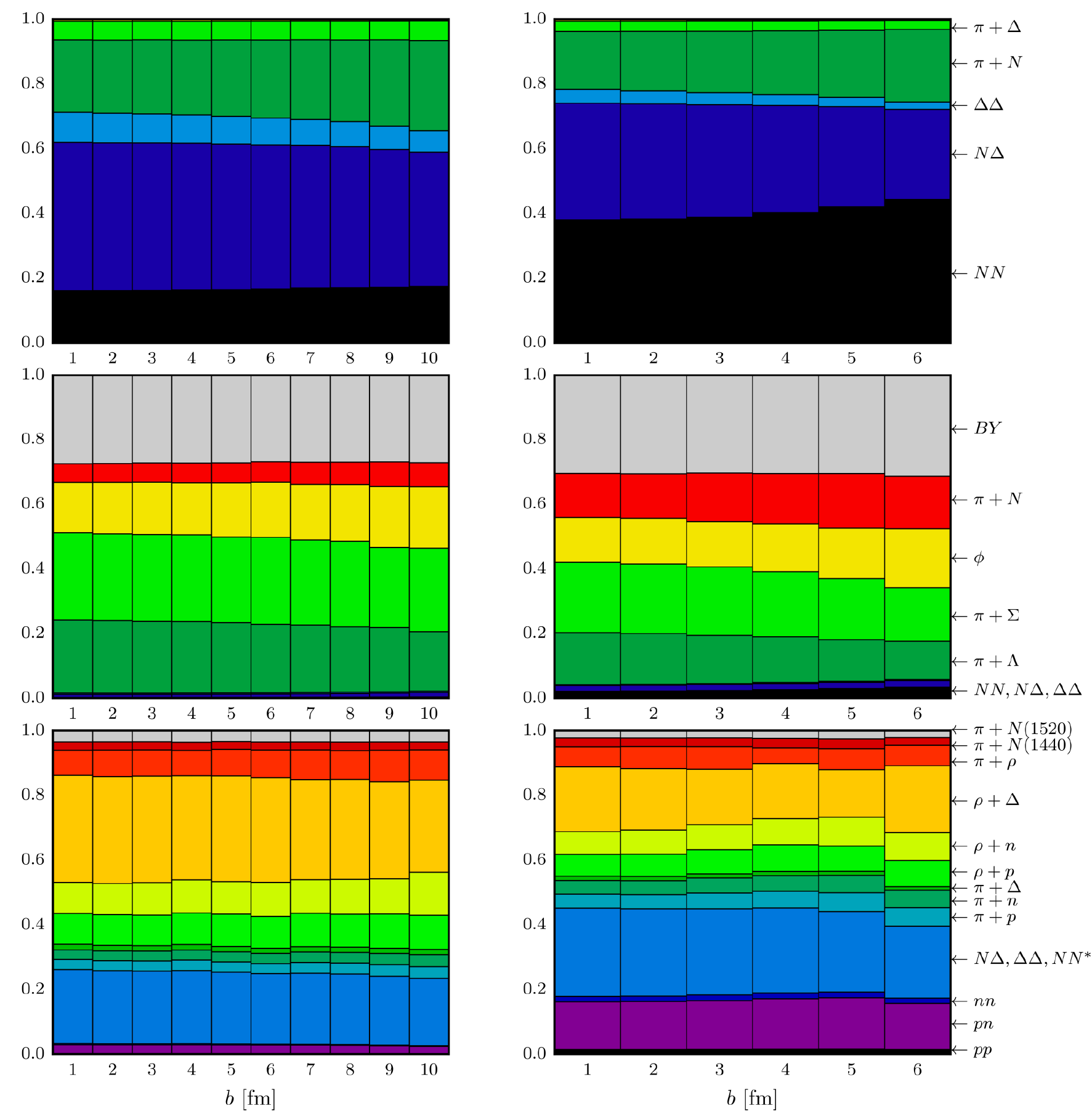}
\vspace*{-9mm}
\caption{Relative contributions of various production channels to \( K^+ \) (upper row), \( K^- \) (middle row) and \(\phi\) multiplicity (lower row) at final run
time \(t = 60\)~fm/c for various impact parameters for collisions Au(1.23 A GeV) + Au (left) and Ar(1.756 A GeV) + KCl (right). See Appendix A of \cite{BRabe}
for details of the reaction channels and cross sections.}
 \label{fig:3.5}
 \end{figure}  

Our interpretation is hampered in some details since we fail to reproduce accurately
the $\phi$ multiplicity and the $K^\pm, \phi$ slope parameters in the
analyzed reaction Au(1.23 A GeV) + Au. The same code, however, describes well
the available data of the reaction Ar(1.756 A GeV) + KCl (see Appendix A). 
It happens  that the
weights of various channels are fairly different, see the panels in Fig.~\ref{fig:3.5}.
This means that one needs to investigate different system sizes and different
beam energies for benchmarking both, the many microscopic input data and
their processing in simulation codes by a limited number of observables.  
In so far, the here presented results can serve as useful reference of
analyses of strangeness dynamics of the planned experiments of CBM.   

In summary we put forward arguments in favor of a dynamical 
(i.e.\ continuous) kinetic
freeze-out of strangeness carrying mesons in sub-threshold heavy-ion collisions.
The in-depth exploration of strangeness dynamics in threshold-near heavy-ion
collisions at SIS18 is a valuable prerequisute for investigating multi-strange
hadron phenomena at higher beam energies, e.g.\ at SIS100, NICA, JPARC
etc.\ The insights gained in the strangeness sector pave the way for 
future analog
studies of charm degrees of freedom as probes of compressed baryon matter.   

{\bf Acknowledgments:}
The authors gratefully acknowledge the collaboration with and within
HADES and CBM, in particular with
R.~Kotte, J.~Stroth, T.~Galatyuk, and M.~Lorenz.  

\begin{appendix}

\section{A\lowercase{r}(1.756 A GeV) + KC\lowercase{l}}
\begin{figure}[htb!]
	\centering
	\begin{minipage}{0.5\textwidth}
		\includegraphics{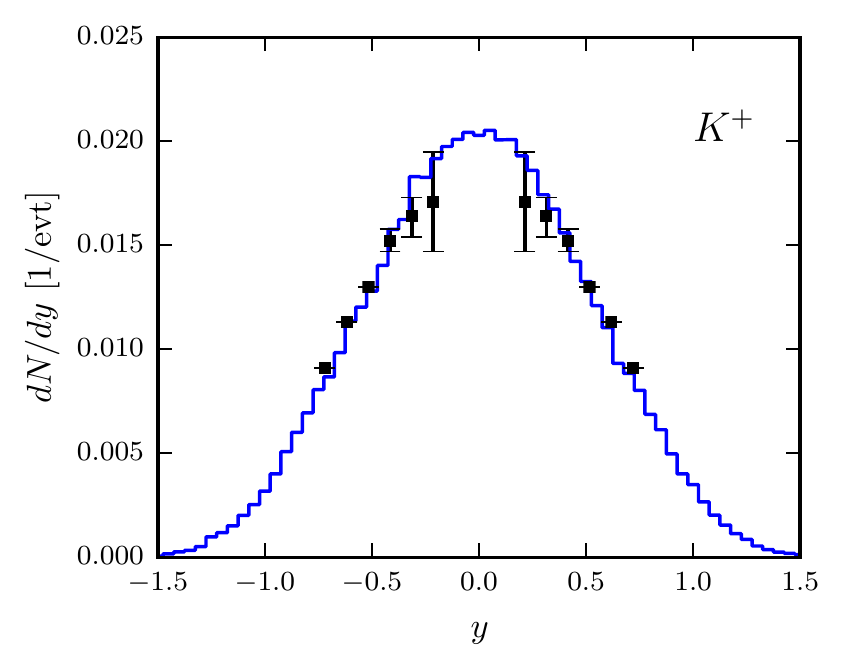}
	\end{minipage}
	\begin{minipage}{0.49\textwidth}
		\includegraphics{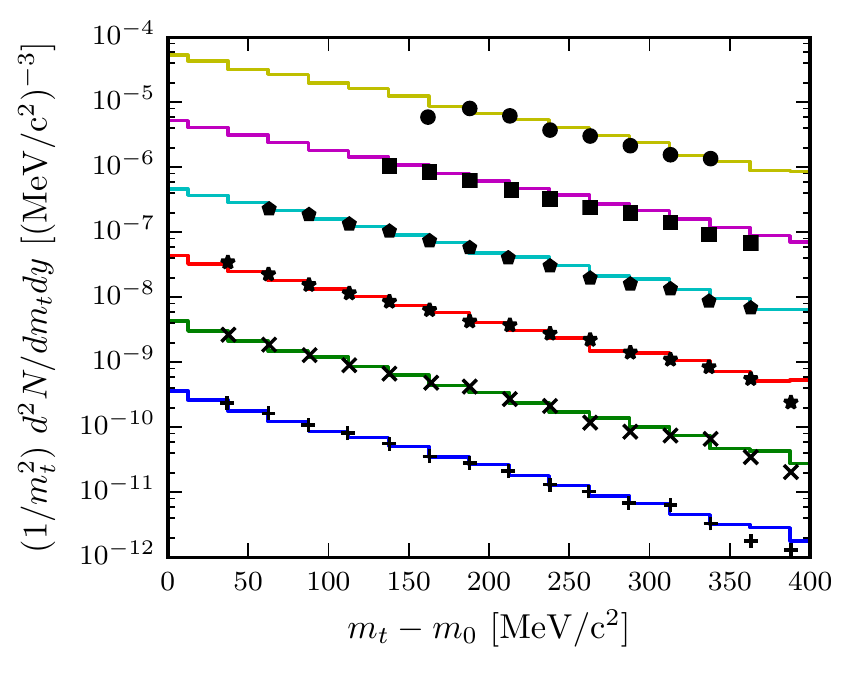}
	\end{minipage}
\vspace*{-3mm}
\caption{Left panel: rapidity spectrum of \( K^+ \) mesons in the center-of-mass system for impact parameter \(b = 3.9\)~fm. Right panel: transverse mass spectra for \(b = 3.9\)~fm for the six rapidity intervals \( 0.1 \leq y_{lab} \leq 0.2 \) to \(0.6 \leq y_{lab} \leq 0.7\) (bottom to top) with scaling factors \(10^0\) to \(10^5\). The black points represent the experimental data from \cite{Agakishiev:2009ar} 
under LVL1 trigger setting.}
 \label{fig:4.2}
 \end{figure}

\begin{figure}[htb!]
\centering
\begin{minipage}{0.51\textwidth}
	\includegraphics{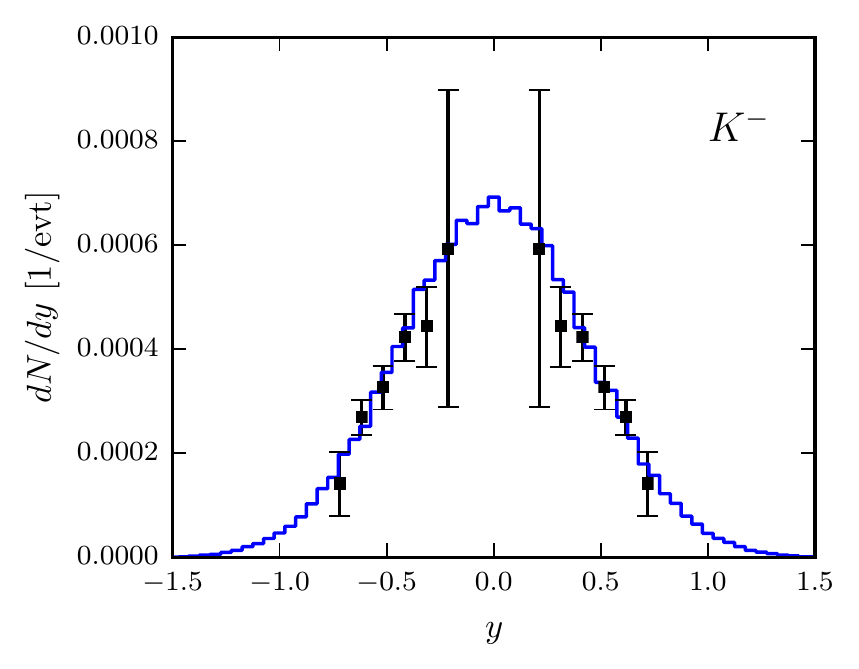}
\end{minipage}
\begin{minipage}{0.48\textwidth}
	\includegraphics{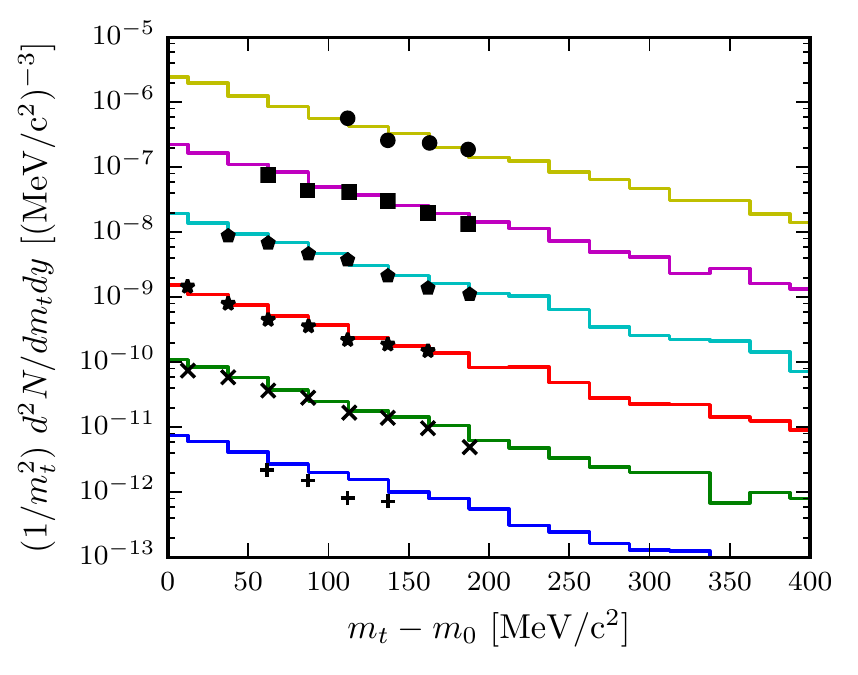}
\end{minipage}
\vspace*{-3mm}
\caption{Same as Fig.~\ref{fig:4.2}, but for \(K^-\) mesons.}
 \label{fig:4.3}
 \end{figure}

\begin{figure}[htb!]
\centering
\begin{minipage}{0.515\textwidth}
	\includegraphics{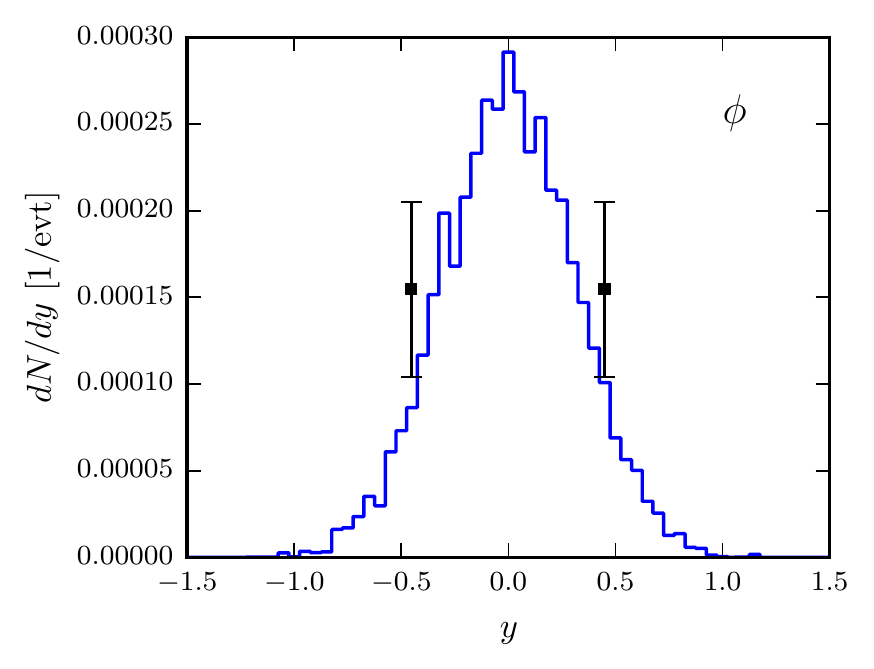}
\end{minipage}
\begin{minipage}{0.475\textwidth}
	\includegraphics{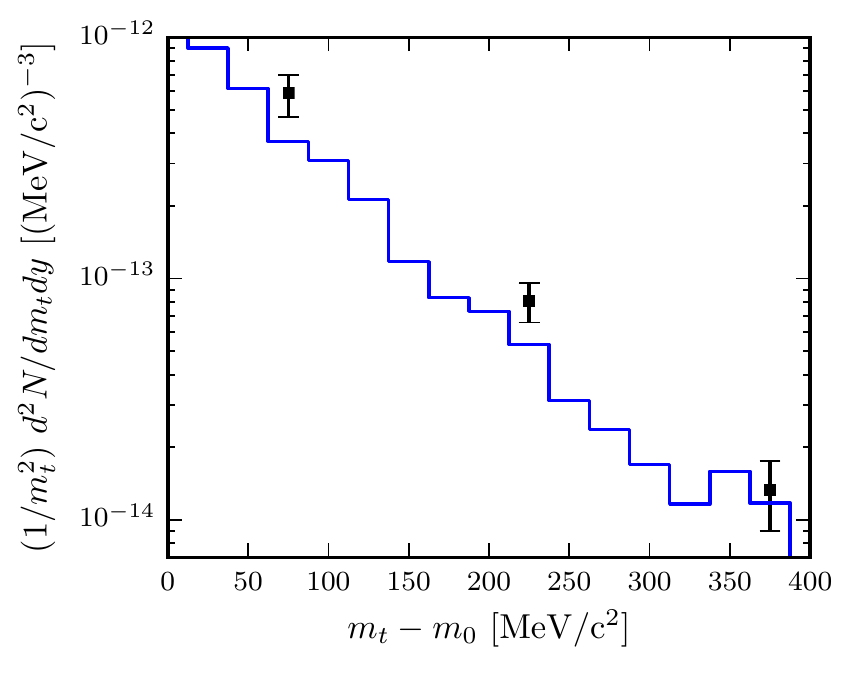}
\end{minipage}
\vspace*{-3mm}
\caption{Left panel: same as Fig.~\ref{fig:4.2}, left panel, but for \(\phi\) mesons. Right panel: transverse mass spectrum for \(b = 3.9\)~fm for the rapidity interval \( 0.2 \leq y_{lab} \leq 0.6 \) and experimental data from \cite{Agakishiev:2009ar} as black points.}
 \label{fig:4.4}
 \end{figure}

\begin{figure}[htb!]
 	\centering
 	\includegraphics{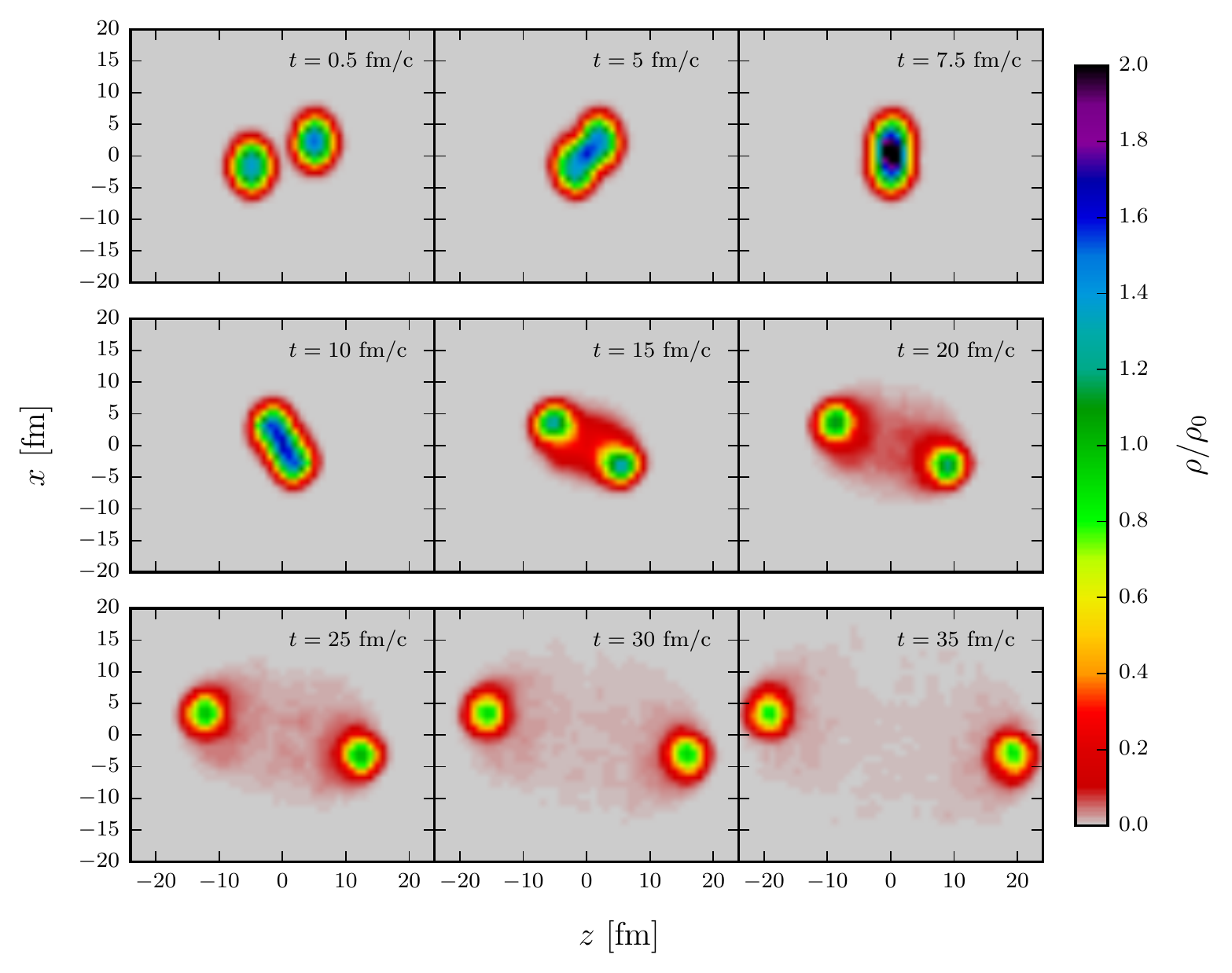}
\vspace*{-3mm}
 	\caption{Same as Figure \ref{fig:3.6}, but for Ar(1.756 A GeV) + KCl and for the impact parameter \( b = 3.9 \)~fm.}
 \label{fig:4.7}
 \end{figure}

\begin{figure}[htb!]
\centering
\includegraphics{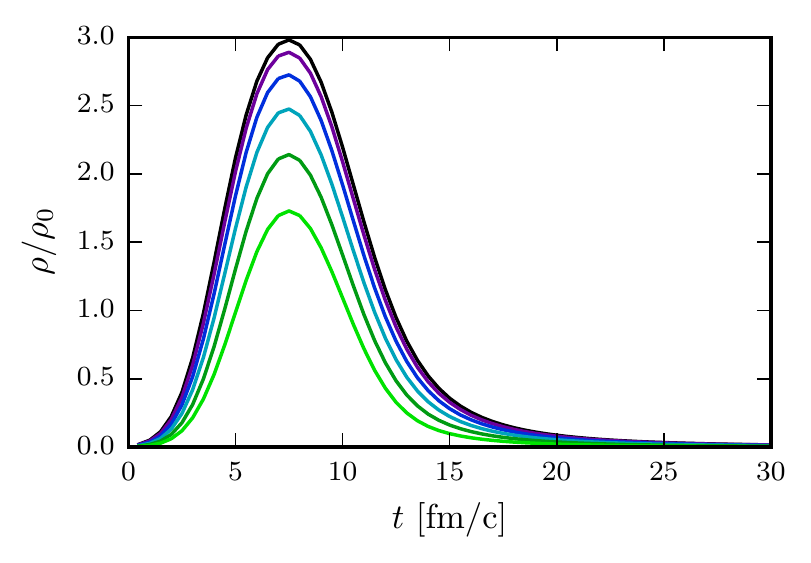}
\vspace*{-3mm}
\caption{Same as Figure \ref{fig:3.7}, but for Ar(1.756 A GeV) + KCl and for the impact parameters \( 1 \leq b \leq 6 \)~fm.}
 \label{fig:4.8}
 \end{figure}

\begin{figure}[htb!]
\centering
\begin{minipage}{0.495\textwidth}
	\includegraphics{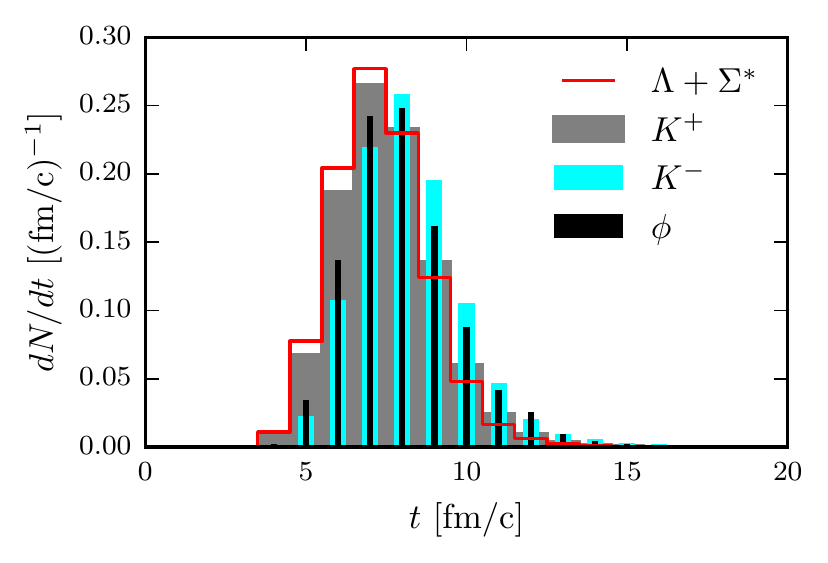}
\end{minipage}
\begin{minipage}{0.495\textwidth}
	\includegraphics{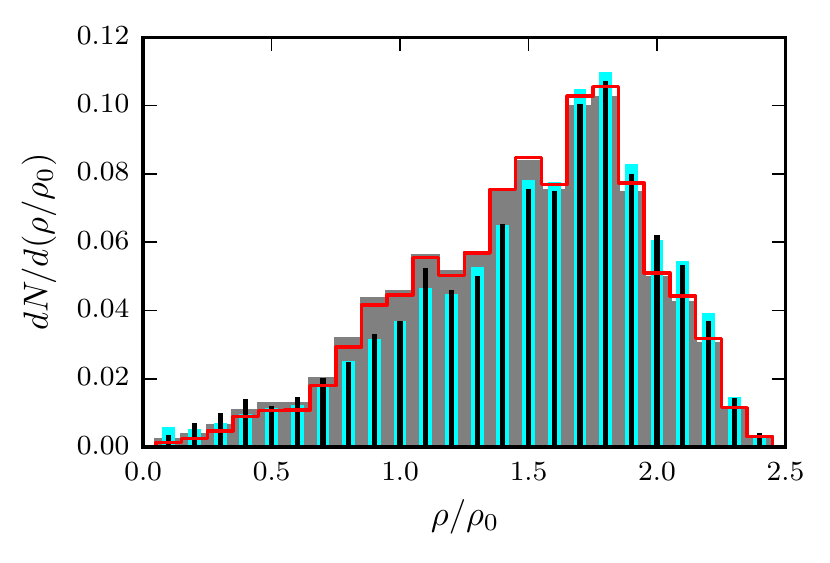}
\end{minipage}
\vspace*{-6mm}
\caption{Same as Figure \ref{fig:3.8}, but for Ar(1.756 A GeV) + KCl and for \(b = 3.9\)~fm.}
 \label{fig:4.9}
 \end{figure}

\begin{figure}[htb!]
\centering
\begin{minipage}{0.495\textwidth}
	\includegraphics{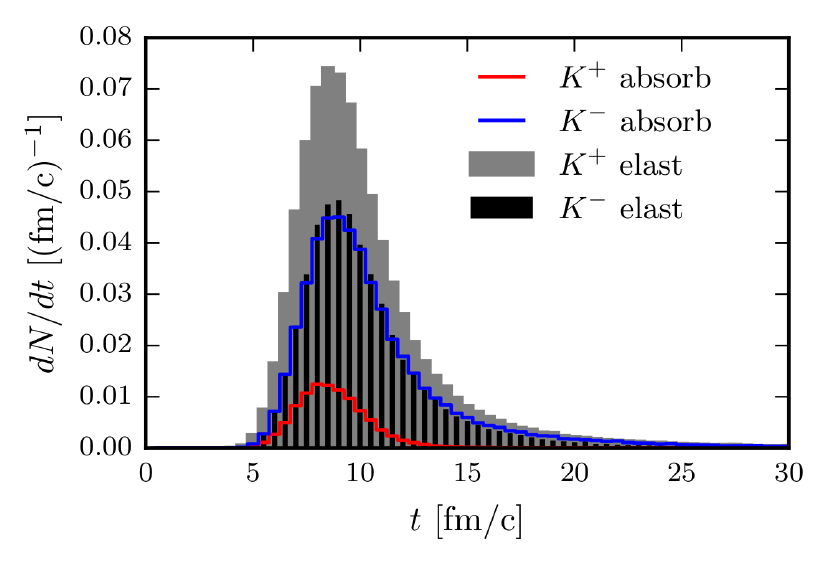}
\end{minipage}
\begin{minipage}{0.495\textwidth}
	\includegraphics{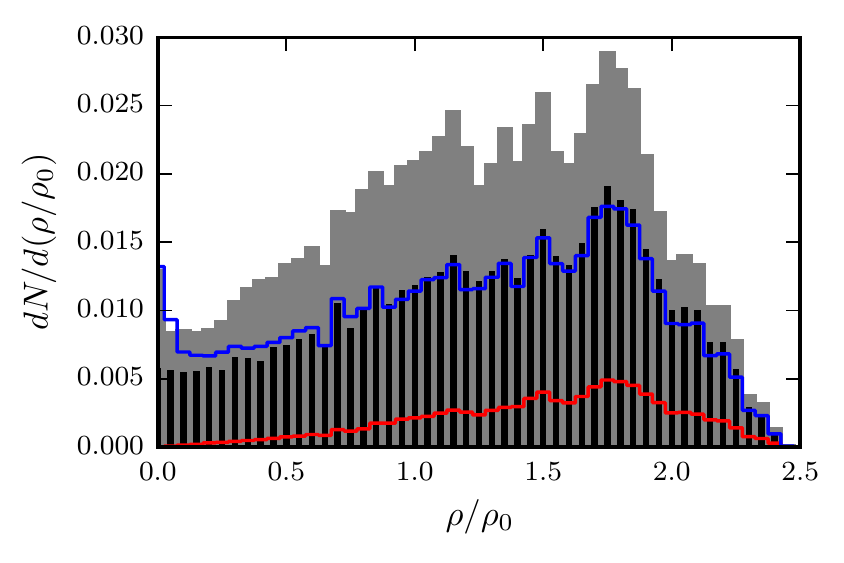}
\end{minipage}
\vspace*{-6mm}
\caption{Same as Figure \ref{fig:3.10}, but for Ar(1.756 A GeV) + KCl and for \(b = 3.9\)~fm.}
 \label{fig:4.11}
 \end{figure}

\begin{figure}[htb!]
\centering
\begin{minipage}{0.495\textwidth}
	\includegraphics{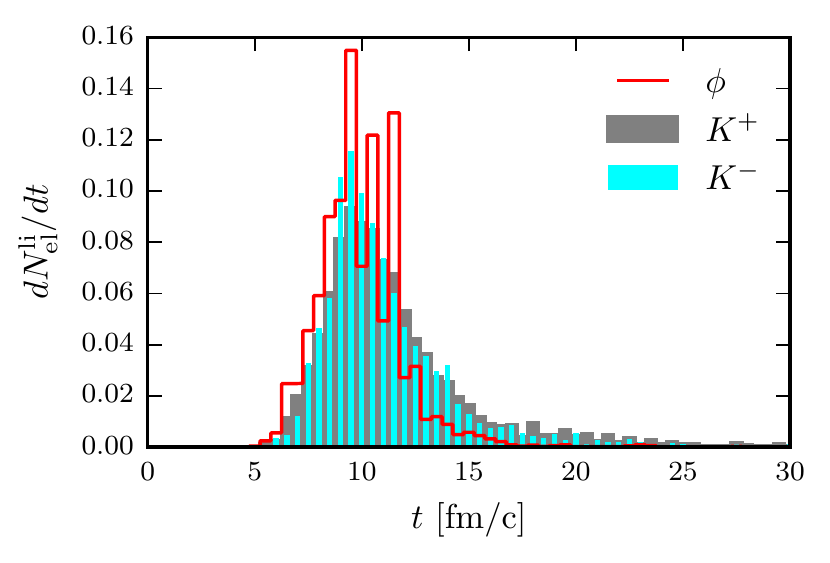}
\end{minipage}
\begin{minipage}{0.495\textwidth}
	\includegraphics{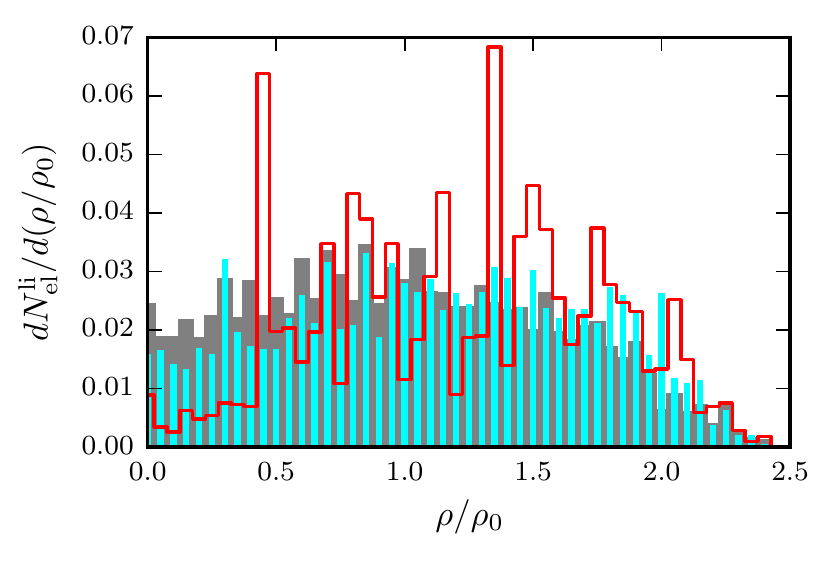}
\end{minipage}
\vspace*{-6mm}
\caption{Same as Figure \ref{fig:3.11}, but for Ar(1.756 A GeV) + KCl 
and for \(b = 3.9\)~fm. Note our restriction to the last {\em elastic} interaction.
In contrast, Fig.~10 in \cite{Schade:2009gg} is based on a counting scheme
of {\em all} last interactions. Due to the partially perturbative treatment
\cite{Schade:2009gg} of $K^-$ absorption in dilute nuclear matter, the
apparently late $K^-$ interaction is overestimated. }
 \label{fig:4.12}
 \end{figure}

\begin{figure}[htb!]
\centering
\includegraphics{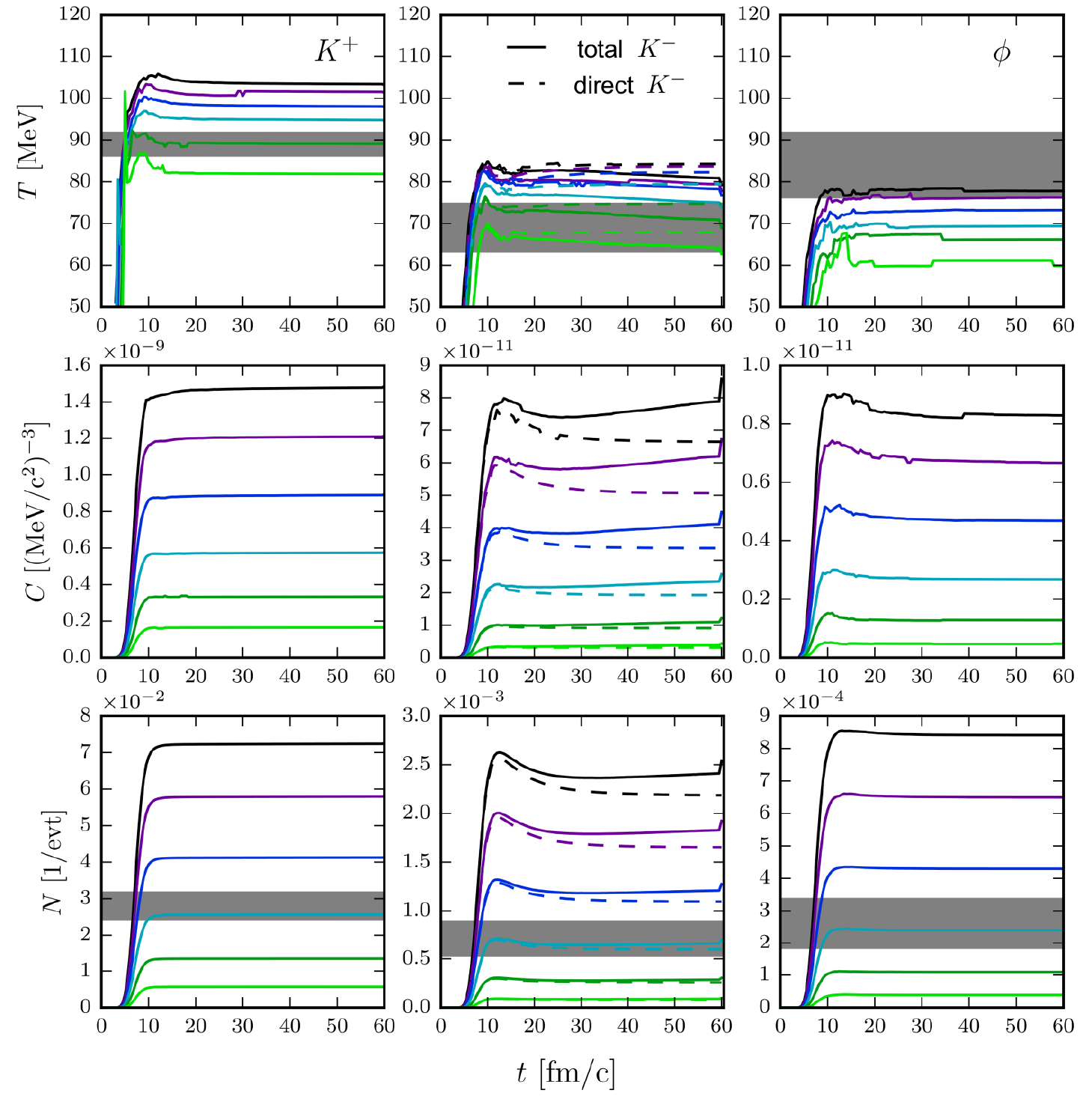}
\vspace*{-3mm}
\caption{Same as Figure \ref{fig:3.12}, but for Ar(1.756 A GeV) + KCl and for impact parameters \(1 \leq b \leq 6\)~fm. The experimental data \cite{Agakishiev:2009ar} are represented as gray bars.}
 \label{fig:4.13}
 \end{figure}

\begin{figure}[ht!]
\centering
\includegraphics{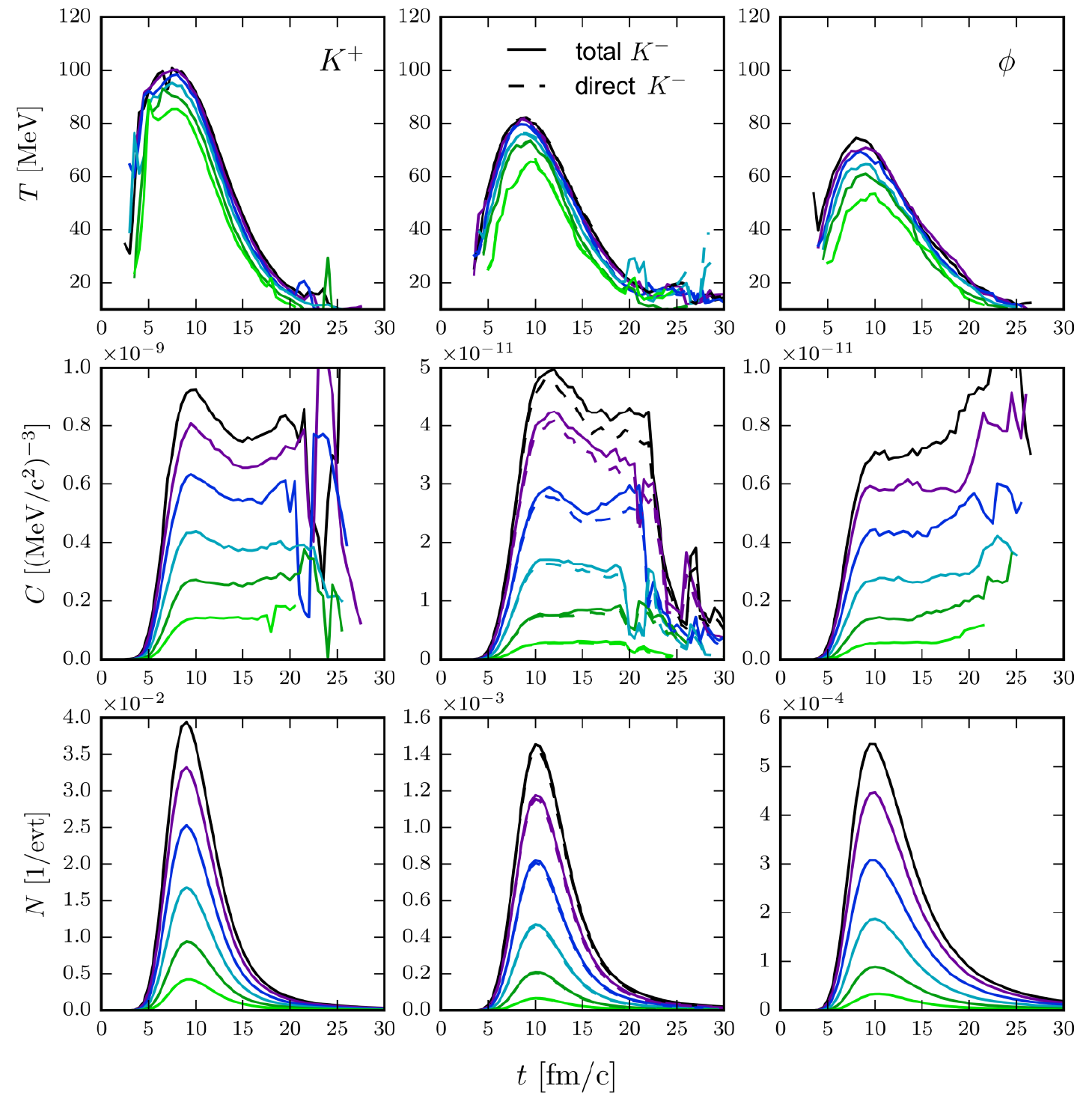}
\vspace*{-6mm}
\caption{Same as Figure \ref{fig:3.14}, but for Ar(1.756 A GeV) + KCl and for impact parameters \(1 \leq b \leq 6\)~fm in steps of  1 fm (from top to bottom).}
 \label{fig:4.15}
 \end{figure}

\begin{figure}[hb!]
\centering
\begin{minipage}{0.495\textwidth}
	\includegraphics{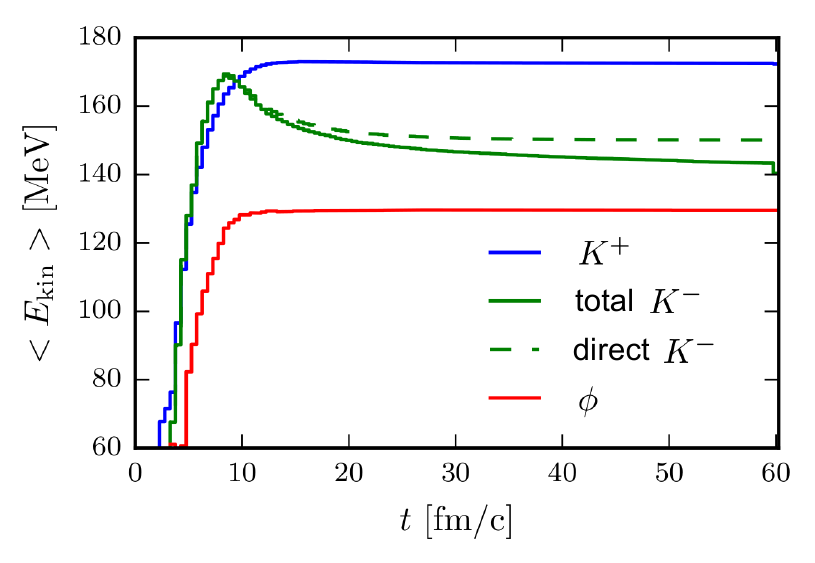}
\end{minipage}
\begin{minipage}{0.495\textwidth}
	\includegraphics{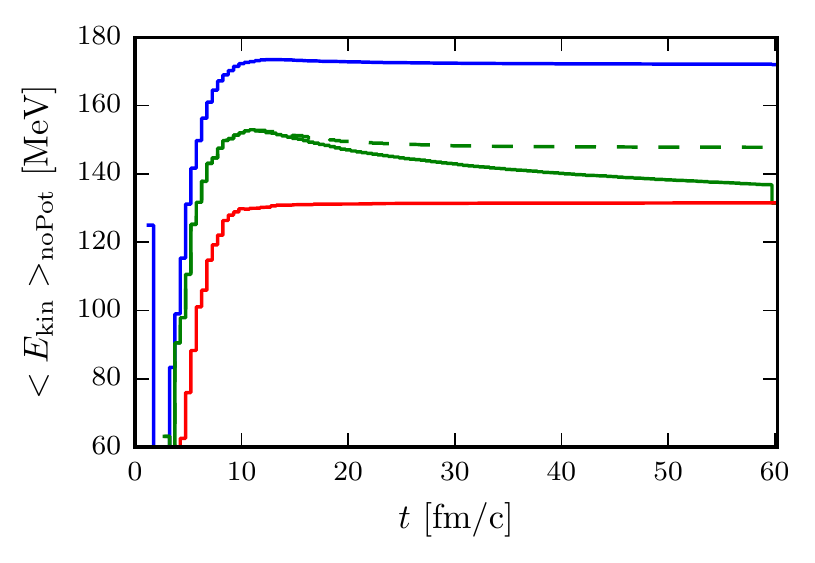}
\end{minipage}
\vspace*{-6mm}
\caption{Same as Figure \ref{fig:3.18}, but for Ar(1.756 A GeV) + KCl and for \(b = 3.9\)~fm.}
 \label{fig:4.18}
 \end{figure}

To enable an easy one-to-one comparison of the results presented in
section III we recollect some analog results for collisions Ar(1.756 A GeV) + KCl
in Figs.~\ref{fig:4.2} - \ref{fig:4.18}. 
Further details can be found in \cite{Schade:2009gg}.
 

Instead of scanning through impact parameters, as done in Figs.~\ref{fig:3.1} -
\ref{fig:3.3}, we compare in Figs.~\ref{fig:4.2} and \ref{fig:4.3} the rapidity 
and transverse momentum spectra with data for one optimized impact parameter,
$b = 3.9$~fm. The latter value corresponds to the mean impact parameter
enforced by the trigger setting LVL1 in the experiment  \cite{Agakishiev:2009ar}.
We emphasize the quite accurate agreement of data and simulations
for $K^\pm$. The $\phi$ slope parameter is acceptable (see right panel
in Fig.~\ref{fig:4.4}, but the yield, to be extracted from the rapidity distribution 
(left panel in Fig.~\ref{fig:4.4}) is only marginally consistent with data.

The high-density stage is much shorter for Ar(1.756 A GeV) + KCl
(see Figs.~\ref{fig:4.7} and \ref{fig:4.8}) than for Au(1.23 A GeV) + Au
(see Figs.~\ref{fig:3.6} and \ref{fig:3.7}). Correspondingly, the various rates
of production / absorption / elastic scattering / last elastic scattering of 
$K^\pm$, $\phi$ are concentrated on shorter time intervals, see
left panels of Figs.~\ref{fig:4.9} - \ref{fig:4.12} and compare with
Figs.~\ref{fig:3.8} - \ref{fig:3.11}. The same rates, however, as a function
of local density, look very similar for the production channels
(compare right panels of Fig.~\ref{fig:4.9} with Fig.~\ref{fig:3.8}),
while the elastic and absorption rates of $K^\pm$ peak at somewhat higher 
local densities, with the exception of $K^+$ absorption (compare right
panels in Fig.~\ref{fig:4.11} with Fig.~\ref{fig:3.10}). The last elastic
$K^\pm,\phi$ scatterings are fairly smoothly distributed over all densities
(see right panel in Fig.~\ref{fig:4.12}, while for Au(1.23 A GeV) + Au
an apparent peaking at $\rho = (1 \cdots 1.5) \rho_0$ occurs (see right panel
in Fig.~\ref{fig:3.11}). Such considerations have the motivation to elucidate
whether the strangeness-carrying probes are specific for certain density ranges.        

The patterns of the time dependence of slope parameters and normalizations
of Ar(1.756 A GeV)+ KCl look very similar to the case of Au(1.23 A GeV) + Au,
where however the experimental values of the former one are nicely reproduced,
see Fig.~\ref{fig:4.13} and compare with Fig.~\ref{fig:3.12}. An analog statement
holds for the slope parameters and normalizations in the central cell - again
within shorter time intervals for Ar(1.756 A GeV) + KCl (see Fig.~\ref{fig:4.13}
and compare with Fig.~\ref{fig:3.14}).  

Comparing the time evolution of the mean kinetic energies of $K^\pm, \phi$ 
with and without $KN$ potentials one observes analog patterns in both
collision systems, see Figs.~\ref{fig:3.18} and \ref{fig:4.18}. However, the local
maximum for $K^-$ and the onset of the flat sections for $K^+$ and $\phi$
are achieved earlier for Ar(1.756 A GeV) + KCl, and the different beam
energies reflect themselves in higher values.
The sharp drops of the $K^-$ mean kinetic momentum (green curves)
at $t = 60$~fm/c are due to the $K^-$ component from $\phi$ decays added
at the end of the simulation run time. This is the "cooling of the $K^-$ 
spectra" by $\phi$ decay \cite{Adamczewski-Musch:2017rtf}. 

\end{appendix}


\clearpage

\begin{thebibliography}{99}

\bibitem{Forster:2003vc} 
  A.~Forster {\it et al.} [KaoS Collaboration],
  ``First evidence for different freezeout conditions for kaons and anti-kaons observed in heavy ion collisions,''
  Phys.\ Rev.\ Lett.\  {\bf 91}, 152301 (2003)
  [nucl-ex/0307017].

\bibitem{Adamczewski-Musch:2017rtf} 
  J.~Adamczewski-Musch {\it et al.} [HADES Collaboration],
  ``Deep sub-threshold $\phi$ production in Au+Au collisions,''
  Phys.\ Lett.\ B {\bf 778}, 403 (2018)
  [arXiv:1703.08418 [nucl-ex]].

\bibitem{Kotte_BK}
R.~Kotte and B.~K\"ampfer, 
"Acceptance and count rate estimates for experiments on sub-threshold
$phi$ meson production in central collisions of C + C at 2 A GeV," 
FZR-339 (2002)

\bibitem{Kampfer:2001mc} 
  B.~Kampfer, R.~Kotte, C.~Hartnack and J.~Aichelin,
  ``Phi puzzle in heavy ion collisions at 2/A-GeV: How many K- from phi decays?,''
  J.\ Phys.\ G {\bf 28}, 2035 (2002)
  [nucl-th/0112040].

\bibitem{Lorenz:2010zz} 
  M.~Lorenz [for the HADES Collaboration],
  ``Investigating dense nuclear matter with rare hadronic probes,''
  PoS BORMIO {\bf 2010}, 038 (2010).

\bibitem{Hartnack:2011cn} 
  C.~Hartnack, H.~Oeschler, Y.~Leifels, E.~L.~Bratkovskaya and J.~Aichelin,
  ``Strangeness Production close to Threshold in Proton-Nucleus and Heavy-Ion Collisions,''
  Phys.\ Rept.\  {\bf 510}, 119 (2012)
  [arXiv:1106.2083 [nucl-th]].

\bibitem{Steinberg:2018jvv} 
  V.~Steinberg, J.~Staudenmaier, D.~Oliinychenko, F.~Li, \"O.~Erkiner and H.~Elfner,
  ``Strangeness production via resonances in heavy-ion collisions at energies available at the GSI Schwerionensynchrotron,''
  Phys.\ Rev.\ C {\bf 99}, no. 6, 064908 (2019)
  [arXiv:1809.03828 [nucl-th]].

\bibitem{Weil:2016zrk} 
  J.~Weil {\it et al.},
  ``Particle production and equilibrium properties within a new hadron transport approach for heavy-ion collisions,''
  Phys.\ Rev.\ C {\bf 94}, no. 5, 054905 (2016)
  [arXiv:1606.06642 [nucl-th]].

\bibitem{Inghirami:2019muf} 
  G.~Inghirami, P.~Hillmann, B.~Tomášik and M.~Bleicher,
  ``Temperatures and chemical potentials at kinetic freeze-out in relativistic heavy ion collisions from coarse grained transport simulations,''
  arXiv:1909.00643 [hep-ph].

\bibitem{Steinheimer:2015sha} 
  J.~Steinheimer and M.~Bleicher,
  ``Sub-threshold $\phi$ and $\Xi^-$ production by high mass resonances with UrQMD,''
  J.\ Phys.\ G {\bf 43}, no. 1, 015104 (2016)
  [arXiv:1503.07305 [nucl-th]].

\bibitem{Kolb_Turner} 
E.~ W.~Kolb and M.~ S.~Turner,
"The Early Universe,"
Front. Phys. 69 (1990) 1  

\bibitem{Andronic:2017pug} 
  A.~Andronic, P.~Braun-Munzinger, K.~Redlich and J.~Stachel,
  ``Decoding the phase structure of QCD via particle production at high energy,''
  Nature {\bf 561}, no. 7723, 321 (2018)
  [arXiv:1710.09425 [nucl-th]].

\bibitem{Agakishiev:2015bwu} 
  G.~Agakishiev {\it et al.} [HADES Collaboration],
  ``Statistical model analysis of hadron yields in proton-nucleus and heavy-ion collisions at SIS 18 energies,''
  Eur.\ Phys.\ J.\ A {\bf 52}, no. 6, 178 (2016)
  [arXiv:1512.07070 [nucl-ex]].

\bibitem{Agakishiev:2010rs} 
  G.~Agakishiev {\it et al.} [HADES Collaboration],
  ``Hyperon production in Ar+KCl collisions at 1.76A GeV,''
  Eur.\ Phys.\ J.\ A {\bf 47}, 21 (2011)
  [arXiv:1010.1675 [nucl-ex]].

\bibitem{Motornenko:2019jha} 
  A.~Motornenko, V.~Vovchenko, C.~Greiner and H.~Stoecker,
  ``Kinetic freeze-out temperature from yields of short-lived resonances,''
  arXiv:1908.11730 [hep-ph].

\bibitem{Kapusta:2018omb} 
  J.~I.~Kapusta and M.~Li,
  ``High Baryon Densities Achieveable at RHIC and LHC,''
  Nucl.\ Phys.\ A {\bf 982}, 903 (2019)
  [arXiv:1807.10823 [nucl-th]].

\bibitem{Friman:2011zz} 
  B.~Friman, C.~Hohne, J.~Knoll, S.~Leupold, J.~Randrup, R.~Rapp and P.~Senger,
  ``The CBM physics book: Compressed baryonic matter in laboratory experiments,''
  Lect.\ Notes Phys.\  {\bf 814}, pp.1 (2011).
  doi:10.1007/978-3-642-13293-3

\bibitem{Ablyazimov:2017guv} 
  T.~Ablyazimov {\it et al.} [CBM Collaboration],
  ``Challenges in QCD matter physics --The scientific programme of the Compressed Baryonic Matter experiment at FAIR,''
  Eur.\ Phys.\ J.\ A {\bf 53}, no. 3, 60 (2017)
  [arXiv:1607.01487 [nucl-ex]].

\bibitem{Schade:2009gg} 
  H.~Schade, G.~Wolf and B.~Kampfer,
  ``Role of phi decays for K- yields in relativistic heavy-ion collisions,''
  Phys.\ Rev.\ C {\bf 81}, 034902 (2010)
  [arXiv:0911.3762 [nucl-th]].

\bibitem{Agakishiev:2009ar} 
  G.~Agakishiev {\it et al.} [HADES Collaboration],
  ``Phi decay: A Relevant source for K- production at SIS energies?,''
  Phys.\ Rev.\ C {\bf 80}, 025209 (2009)
  [arXiv:0902.3487 [nucl-ex]].

\bibitem{Kolomeitsev:2004np} 
  E.~E.~Kolomeitsev {\it et al.},
  ``Transport theories for heavy ion collisions in the 1-A-GeV regime,''
  J.\ Phys.\ G {\bf 31}, S741 (2005)
  [nucl-th/0412037].

\bibitem{Adamczewski-Musch:2017sdk} 
  J.~Adamczewski-Musch {\it et al.} [HADES Collaboration],
  ``Centrality determination of Au + Au collisions at 1.23A GeV with HADES,''
  Eur.\ Phys.\ J.\ A {\bf 54}, no. 5, 85 (2018)
  [arXiv:1712.07993 [nucl-ex]].

\bibitem{BRabe}
B.~Rabe,
"Untersuchung der Ausfrier-Dynamik von Kaonen in relativistischen
Schwerionenkollisionen,"
Master Thesis, TU Dresden (2019)

\bibitem{Gasik:2015zwm}
P.~Gasik {\it et al.} [FOPI Collaboration],
``Strange meson production in Al+Al collisions at 1.9 A GeV,''
Eur.\ Phys.\ J.\ A {\bf 52} (2016) no.6,  177
[arXiv:1512.06988 [nucl-ex]].

\bibitem{Piasecki:2018psj}
K.~Piasecki {\it et al.} [FOPI Collaboration],
``Wide acceptance measurement of the K$^-$/K$^+$ ratio from Ni+Ni collisions at 1.91A GeV,''
Phys.\ Rev.\ C {\bf 99} (2019) no.1,  014904
[arXiv:1807.00576 [nucl-ex]].


\end{thebibliography}
\end{document}